\documentclass[manuscript,screen]{acmart}
\usepackage{graphicx}
\usepackage{comment}
\usepackage{natbib}
\usepackage{multirow} 
\usepackage{lineno,hyperref}

\AtBeginDocument{%
  }

\setcopyright{acmlicensed}
\copyrightyear{2025}
\acmYear{2025}
\acmDOI{XXXXXXX.XXXXXXX}

\acmJournal{TOSEM}
\begin{document}

\title{Towards an Understanding of Developer Experience-Driven Transparency in Software Ecosystems}

\author{Rodrigo Oliveira Zacarias}
\email{rodrigo.zacarias@edu.unirio.br}
\orcid{0000-0003-0005-4669}
\affiliation{%
  \institution{Federal University of the State of Rio de Janeiro (UNIRIO)}
  \city{Rio de Janeiro}
  \country{Brazil}
}

\author{Rodrigo Pereira dos Santos}
\email{rps@uniriotec.br}
\orcid{0000-0003-4749-2551}
\affiliation{%
  \institution{Federal University of the State of Rio de Janeiro (UNIRIO)}
  \city{Rio de Janeiro}
  \country{Brazil}}

\author{Patricia Lago}
\email{p.lago@vu.nl}
\orcid{0000-0002-2234-0845}
\affiliation{%
  \institution{Vrije Universiteit Amsterdam}
  \city{Amsterdam}
  \country{The Netherlands}
}

\renewcommand{\shortauthors}{Zacarias et al.}

\begin{abstract}
Software ecosystems (SECO) have become a dominant paradigm in the software industry, where third-party developers co-create value through complementary components and services. While developer experience (DX) is increasingly recognized as critical for a successful and sustainable SECO, transparency remains an underexplored factor shaping how developers interact with and perceive ecosystems. Existing research acknowledges transparency as a conditioning element for trust, fairness, and engagement, yet its relationship with DX has not been systematically characterized. Hence, this work aims to advance the understanding of transparency in SECO from a developer-centered perspective. To this end, we propose SECO-TransDX (\textit{Transparency in Software Ecosystems from a Developer Experience Perspective}), a conceptual model that introduces the notion of DX-driven transparency. The model identifies key concepts, conditioning factors, common ecosystem procedures, artifacts, and relational dynamics that shape how transparency is perceived and constructed from the developer’s perspective during interactions with SECO. SECO-TransDX was built upon insights from prior studies on conditioning factors for transparency, DX factors, and the SECO meta-model, and was further evaluated and refined through a Delphi study with experts from academia and industry. The resulting model consolidates 63 concepts and their relationships, providing a structured lens to examine transparency from a DX perspective in SECO. It highlights five main findings that clarify how transparency mediates the DX across technical, social, and organizational dimensions. Beyond its theoretical contributions, SECO-TransDX also has implications for both academia and industry. For researchers, it offers a structured lens to analyze how transparency mediates DX and provides a foundation for developing guidelines, evaluation frameworks, and practical tools to support more transparent and developer-centered platforms. For practitioners, the model illustrates how different factors and interactions influence developers’ perception of transparency, providing a basis for designing transparent practices and platforms that pave the way for more successful and trustworthy SECO. 
\end{abstract}




\maketitle

\section{Introduction}\label{sec:intro}

Software ecosystems (SECO) have emerged as a dominant paradigm in the software industry, in which a common technological platform is extended by a community of third-party developers who co-create value through complementary software components, services, and applications~\cite{Jansen2020}. Such a common technological platform provides a shared set of resources, including documentation, source code, application programming interfaces (API), software development kits (SDK), and communication channels, to support development and integration efforts~\cite{Manikas2016, meireles2019instrumento}. This platform-centric openness fosters innovation at scale, accelerates platform evolution, and generates mutual benefits for central organizations and third-party developers~\cite{Manikas2016}. Prominent examples of such platform-centered SECO include Android\footnote{\url{https://developer.android.com/}}, iOS\footnote{\url{https://developer.apple.com/}}, GitHub\footnote{\url{https://github.com/}}, and HarmonyOS\footnote{\url{https://developer.harmonyos.com/}}.

As SECO evolve, the experience of developers interacting with platforms has become a critical concern for both researchers and practitioners. The notion of Developer Experience (DX) has emerged as a multidimensional construct that captures how developers feel, think, and act when engaging with software platforms and tools~\cite{Fagerholm2012, Greiler2022}. In a recent study, Zacarias et al.~\cite{zacarias2025Dxfactors} identified a set of core DX factors in the context of SECO, through a systematic mapping study (SMS) and a Delphi study. These factors are organized into four categories: Common Technological Platform, Projects and Applications, Community Interaction, and Expectations and Value of Contribution. They encompass both technical and social aspects of DX and emphasize the need for central organizations to consider the developer perspective when designing and evolving platform interactions. A positive DX is particularly critical during onboarding,  as it shapes whether developers feel motivated, supported, and capable of contributing to the ecosystem. As platforms increasingly compete for users and contributors, delivering a frictionless, developer-centered experience has become a strategic differentiator for a successful and sustainable SECO~\cite{fontao2021developer}.

Among the many factors that shape DX, transparency plays a fundamental yet underexplored role in the context of SECO~\cite{Jansen2013, SantosEtAl2016, meireles2019instrumento}. Transparency enables developers to access and interpret critical artifacts and aspects of an ecosystem, such as software components, governance processes, architectural decisions, contribution dynamics, data practices, and platform evolution\cite{Cataldo2010, Zacarias2024}. By enabling access to these elements, transparency fosters trust, openness, and perceived fairness, factors that influence developers’ motivation to engage and contribute~\cite{hou_systematic_2022}. A lack of transparency can lead to confusion, misalignment, or distrust, discouraging participation and weakening long-term commitment. In contrast, ecosystems that actively promote transparency tend to build stronger community bonds, support more predictable collaboration, and encourage sustained contributor engagement, which are essential for long-term sustainability, i.e., their capacity to remain viable, adaptive, and valuable over time, ensuring continued participation, evolution, and balance among actors~\cite{Zacarias2024sesos}.

Although transparency is conceptually related to DX constructs such as trust, feedback, and findability~\cite{hou_systematic_2022}, it is rarely treated as a first-class concept in SECO research. While previous studies have identified conditioning factors that influence transparency in SECO~\cite{Zacarias2024}, the relationship between DX and transparency remains theoretically unmodeled and empirically unexplored. Addressing this gap requires a clear conceptualization of how transparency is shaped through the lens of the developer. Transparency, while technical, plays a mediating role in shaping developers’ social experiences within the ecosystem. These aspects are central to developer engagement and sustained contribution, which affect the long-term sustainability and business value of SECO. We argue that establishing a shared understanding of DX-driven transparency could advance the research in this field. To this end, this work addresses the following research question (RQ): \textit{``How can transparency in SECO be characterized from a DX perspective?''}

Hence, this work aims to advance a developer-centered understanding of transparency in SECO. Instead of approaching transparency solely as a platform-defined attribute, we explore how it can be characterized based on how developers experience their interactions within the ecosystem. To this end, we propose a conceptual model, named SECO-TransDX (\textit{Transparency in Software Ecosystems from a Developer Experience Perspective}), which introduces the notion of DX-driven transparency. The model identifies key concepts, conditioning factors, common ecosystem procedures, artifacts, and relational dynamics that shape how transparency is perceived and constructed from the developer’s perspective during interactions with SECO.

To build SECO-TransDX, we drew on insights from three prior studies. First, we analyzed results from an SMS and field study to characterize conditioning factors for transparency in SECO, including organizational, technical, and developer-related elements~\cite{Zacarias2024}. Second, we examined findings from an SMS and Delphi study that investigated DX factors influencing third-party developers’ decisions to engage with and continue contributing to SECO~\cite{zacarias2025Dxfactors}. Third, we analyzed the SECO meta-model proposed by Wouters et al.~\cite{Wouters2019}, which identifies key ecosystem entities and their relationships across five thematic areas. Therefore, SECO-TransDX integrates insights from these prior studies and extends them by providing a conceptual model comprising 63 concepts and their relationships. Additionally, SECO-TransDX has been evaluated through a Delphi study, which is an expert judgment method used to assess the model’s relevance and coherence from the perspective of experienced practitioners and researchers~\cite{DALKEY1969408}. 

The main contribution of this work is the introduction of SECO-TransDX as a conceptual model, evaluated by experts through a Delphi study, that offers a shared understanding of transparency in SECO grounded in the literature and software industry practices. The model introduces and formalizes the notion of DX-driven transparency, emphasizing how transparency is perceived, shaped, and constructed through developers’ interactions within the ecosystem. It highlights five main findings that clarify how transparency mediates the DX across technical, social, and organizational dimensions. A conceptual model can be seen as a collection of statements, often represented via a graphical model, that specifies a given domain and helps abstract, understand, and communicate information about it~\cite{Chazette2021}. According to Oliveira Júnior et al.~\cite{OLIVEIRAJR2020301014}, a conceptual model serves as a communication tool to represent concepts and their relationships, especially in emerging and complex domains such as SECO. 

Beyond its theoretical contributions, SECO-TransDX also has implications for both academia and industry. For researchers, it offers a structured lens to analyze how transparency mediates DX and provides a foundation for developing guidelines, evaluation frameworks, and practical tools to support more transparent and developer-centered platforms. For practitioners, the model illustrates how different factors and interactions influence developers’ perception of transparency, providing a basis for designing transparent practices and platforms that pave the way for more sustainable and trustworthy SECO. 

The remainder of this article is organized as follows: Section~\ref{sec:background} presents the background and related work; Section~\ref{sec:method} describes the research method; Section~\ref{sec:model} presents SECO-TransDX; Section~\ref{sec:delphi} reports the evaluation of SECO-TransDX; Section~\ref{sec:discussion} depicts our contributions and main findings, and perspectives for using; Section~\ref{sec:threats} discusses the threats to the validity; and Section~\ref{sec:conclusion} concludes this work e presents future work.

\section{Background}\label{sec:background}
This section presents key concepts related to SECO, as well as definitions of transparency and DX within this context. It sets the foundation for understanding how these elements interact and frames the theoretical gap addressed in this work.

\subsection{Developer Experience and Transparency in Software Ecosystems}\label{sec:secoTransDXBack}
A SECO can be described as a collective of actors and their interactions, operating as a cohesive unit within a distributed market for software products and services. These interactions typically rely on a shared technological platform or marketplace and are facilitated through the exchange of information, resources, and artifacts~\cite{Jansen2009}. To support effective communication and collaboration among developers and users, the integration of supporting tools and mechanisms is essential~\cite{santos2016, Jansen2020}.

Hanssen~\cite{Hanssen2012} identifies three primary roles within SECO based on the nature of their relationship with the platform: (i) keystone, an entity or group responsible for leading platform development; (ii) end-users, who utilize the platform to conduct their business; and (iii) third-party developers, who build complementary solutions or services leveraging the platform. Notably, the platform provider may be represented by a single organization or an open-source community~\cite{Jansen2013}.

Manikas~\cite{Manikas2016} categorizes SECO into three types: proprietary, open-source, and hybrid. Proprietary SECO (PSECO) focus on value creation through proprietary contributions protected by confidentiality agreements (e.g., Microsoft Windows\footnote{\url{https://www.microsoft.com/windows/}} and SAP – Systems, Applications, and Products\footnote{\url{https://www.sap.com/}})~\cite{Outao2025}. Open-source SECO (OSSECO) encourage contributions from diverse actors and communities, often motivated by factors beyond direct financial gain (e.g., Linux Kernel\footnote{\url{https://www.kernel.org/}}, Eclipse Foundation\footnote{\url{https://www.eclipse.org/org/foundation/}}, and Kubernetes\footnote{\url{https://kubernetes.io/}})~\cite{FRANCOBEDOYA2017160}. Hybrid SECO combine both proprietary and open-source approaches, blending platform-led strategies with community-driven contributions, as exemplified by platforms such as Android\footnote{\url{https://developer.android.com/}} and iOS\footnote{\url{https://developer.apple.com/}}~\cite{Fontao2015}.

As SECO grow in complexity and scale, the interactions between developers and the platform become increasingly central to the ecosystem's sustainability. In this context, the notion of DX has emerged as a key lens for understanding how third-party developers perceive, navigate, and engage with SECO. Greiler et al.~\cite{Greiler2022} define DX as ``how developers think about, feel about, and value their work''. This definition builds upon the conceptual framework proposed by Fagerholm and Münch~\cite{Fagerholm2012}, which is grounded in the theory of the trilogy of mind from social psychology~\cite{Hilgard1980}. According to this theory, human experience is composed of three interrelated dimensions: cognition, emotion, and conation. Drawing from this perspective, Fagerholm and Münch~\cite{Fagerholm2012} developed a framework that characterizes DX as a construct shaped by these three psychological dimensions.

The cognitive dimension encompasses the intellectual aspects of how developers interpret and interact with their development environment, including tools, processes, technical skills, and infrastructure. The affective dimension refers to emotional responses related to the development experience, such as feelings of respect, belonging, and social connection within the team or community. The conative dimension involves the developers’ motivations, intentions, and sense of purpose, reflecting how they perceive the value and alignment of their contributions with personal and project goals. 

In SECO, these dimensions are particularly relevant, as developers often interact with decentralized platforms and communities. A deeper understanding of these aspects can help improve engagement and collaboration, especially when attracting and onboarding third-party developers. Fontão et al.~\cite{FontaoEtAl2017} emphasize that DX significantly impacts developer productivity, particularly when publishing applications within a SECO. They emphasize that sustaining ecosystem growth relies on attracting and engaging third-party developers, which requires understanding developers’ expectations and experiences, particularly during their initial interactions with the platform.

Building on our previous work~\cite{zacarias2025Dxfactors}, we consider 27 DX factors that may influence developer engagement in SECO. These are grouped into four categories that capture distinct aspects of developer interaction with the ecosystem: (i) \textit{Common Technological Platform}, which covers infrastructure, tools, documentation, and platform openness; (ii) \textit{Projects and Applications}, focused on development requirements, distribution strategies, and learning curve; (iii) \textit{Community Interaction}, reflecting social dynamics such as recognition, relationship-building, and community support; and (iv) \textit{Expectations and Value of Contribution}, encompassing motivations such as financial return, autonomy, skill development, and long-term engagement. While not exhaustive, this overview provides a conceptual structure for understanding how developers perceive and experience SECO participation.

While these DX factors provide a structured view of what developers value in their interactions with SECO, an underlying aspect that permeates several of these categories is transparency. Recognized as a critical non-functional requirement in software-intensive environments, transparency enables stakeholders to access, interpret, and trust the data and processes managed by software systems~\cite{LeiteCappelli2010, Hosseini2016}. Within SECO, transparency plays a vital role in supporting the engagement and coordination of heterogeneous actors by clarifying how the platform operates and how decisions are made~\cite{Cataldo2010}. Keystone organizations are responsible for fostering this transparency by ensuring the availability of relevant information and facilitating collaboration with third-party developers.

Transparency in SECO spans multiple ecosystem elements, including development processes, platform documentation, governance structures, and the evolution of projects. It supports coordination by making ecosystem activities observable and traceable across organizational boundaries. However, transparency is not inherent to digital environments — it must be deliberately designed and maintained to expose the elements and information that actors consider essential~\cite{SantosEtAl2016}.

In previous work~\cite{Zacarias2024}, we identified a set of conditioning factors (CF) that shape how transparency is perceived and enacted in SECO. Communication channels (CF1) are crucial for facilitating interaction and alignment between actors and the keystone. Accessibility (CF2) and understanding (CF3) of information ensure that all participants can effectively navigate the platform. The quality of platform content (CF4) and the usability of its interfaces (CF5) enhance clarity and reduce friction in accessing information. Auditability (CF6) enables verification of processes and data, while visualization of ecosystem evolution (CF7) improves situational awareness. Finally, the reliability of the information provided by the keystone (CF8) is foundational for building trust and sustaining developer engagement. These findings underscore the relevance of transparency as a key element shaping DX in SECO. However, despite its importance, existing conceptual models in SECO research rarely treat transparency, particularly from the perspective of developers, as a central construct. In the next section, we review related work to further examine this theoretical and practical gap.

\subsection{Theoretical Gap and Related Work}\label{sec:relatedWork}

The concepts of transparency and DX are crucial for the long-term success and sustainability of SECO. Transparency promotes openness, trust, and coordination by making information, processes, and decision-making visible to all stakeholders~\cite{Cataldo2010, Zacarias2024}. DX, in turn, encompasses developers’ perceptions, emotions, motivations, and satisfaction as they engage with the ecosystem~\cite{Fagerholm2012, Greiler2022}. While transparency is primarily a technical and organizational attribute, it directly influences the quality of the DX, which is inherently social and psychological.

Despite their importance, the literature often addresses transparency and DX in isolation. Studies on transparency tend to focus on organizational structures or process visibility without exploring how these practices shape developers’ day-to-day experiences. Conversely, DX research emphasizes cognitive, emotional, and motivational dimensions, but rarely considers how transparency, or lack thereof, contributes to developer satisfaction, engagement, or retention. This conceptual separation limits a holistic understanding of how transparency mechanisms function as enablers of positive developer experiences in SECO. Without an integrated view, ecosystem designers may fail to recognize how technical transparency practices influence social developer dynamics and, ultimately, ecosystem sustainability.

In this work, we aim to bridge this theoretical gap by proposing a conceptual model (SECO-TransDX) that integrates transparency and DX within SECO. The model introduces the notion of DX-driven transparency, highlighting how transparency is experienced, shaped, and constructed through developers’ interactions with the platform. By connecting technical transparency mechanisms with developers’ social and psychological responses, the model offers a comprehensive perspective on how transparency affects DX. To contextualize this gap, we now review key conceptual models addressing SECO from various perspectives. While these models provide valuable insights into SECO structure, governance, and participant dynamics, none explicitly link transparency and DX within a unified framework.

Wouters et al.~\cite{Wouters2019} propose a SECO meta-model that establishes a shared vocabulary for the research domain. Their model standardizes core concepts such as actors, roles, resources, and interactions, offering a structural and organizational view. However, it does not account for how developers experience transparency, nor how it affects engagement and retention. Our model builds upon this foundation by integrating transparency and DX as interrelated constructs, clarifying how transparency practices impact developers’ interactions with the platform.

Oliveira et al.~\cite{Oliveira2020} present a conceptual model focused on SECO governance, highlighting decision-making structures and actor roles. While the authors emphasize communication and transparency as mechanisms for fostering trust, transparency is approached mainly from a governance perspective. Their model does not examine how transparency shapes DX. By contrast, our work proposes a developer-centered model of transparency that incorporates technical, social, and motivational aspects affecting DX.

Malcher et al.~\cite{MALCHER2025107672} explore requirements management in SECO, analyzing coordination challenges and stakeholder alignment. Although their work addresses communication and negotiation, it does not explore how transparency in these procedures influences developers’ experiences and sustained participation. Our model complements their findings by positioning transparency as a technical condition that mediates DX and shapes long-term developer engagement, a critical factor for SECO evolution and success.

In summary, SECO-TransDX distinguishes itself from prior conceptual models by providing a unified representation of transparency and DX as interdependent constructs. Rather than treating transparency as a static platform attribute, the model frames it as a dynamic element that actively shapes how developers think, feel, and act within the ecosystem. This developer-centered perspective is operationalized through a set of well-defined concepts, conditioning factors, common ecosystem procedures, and artifacts that clarify how transparency affects participation and satisfaction. By doing so, SECO-TransDX supports the design of platforms that are not only open and trustworthy, but also more attractive and sustainable from the developer’s point of view.

\section{Research method}\label{sec:method}
\begin{figure}[!ht]
\includegraphics[width=0.7\linewidth]{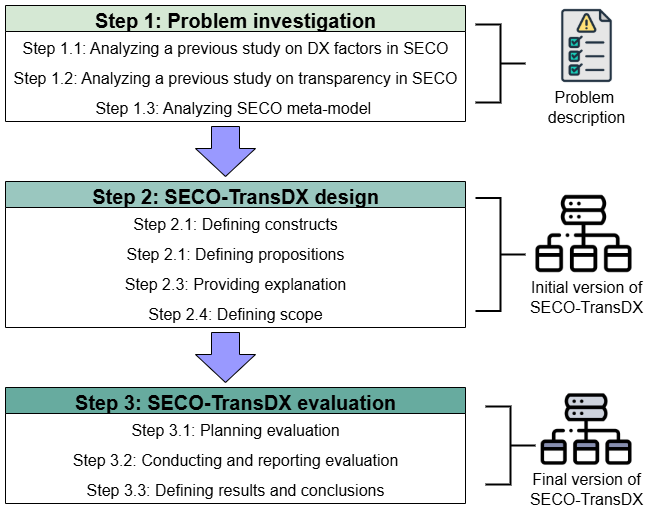}
\caption{Our research method.}
\label{fig:researchMethod}
\Description{Method.}
\end{figure}

To develop the SECO-TransDX conceptual model, we adopted the Design Science Research (DSR) method~\cite{Wieringa2014}, a well-established approach in software engineering and information systems that supports the systematic design, evaluation, and refinement of innovative artifacts. This process was guided by our RQ: \textit{``How can transparency in SECO be characterized from a DX perspective?''} Our research method was structured into three main stages, inspired by the methodological framework proposed by Malcher et al.~\cite{MALCHER2025107672}, as illustrated in Fig.~\ref{fig:researchMethod}.

\subsection{Step 1: Problem investigation}\label{sec:problem}
This first step focused on understanding the research problem by leveraging existing evidence and empirical data from our prior research. It comprised three sub-steps: (1.1) Analyzing a previous study on DX factors in SECO; (1.2) Analyzing a previous study on transparency in SECO; and (1.3) Analyzing the SECO meta-model proposed by Wouters et al.~\cite{Wouters2019} to support the identification of key SECO concepts and relationships. The following sections describe each sub-step in detail.

\subsubsection{Step 1.1: Analyzing a previous study on DX factors in SECO}
In this step, we focused on identifying and understanding the DX factors that influence third-party developers’ decisions to adopt and continue contributing to SECO. To this end, we formulated the following RQ: \textit{``How do DX factors influence third-party developers to adopt and keep contributing to a SECO?''} 

To answer this question, we analyzed results from two prior studies. First, an SMS that reviewed 29 primary studies to assess the state-of-the-art on DX in SECO, identifying 27 factors that affect developer engagement and retention. Second, a Delphi study with 21 third-party developers evaluated the practical influence of these factors. The resulting list of DX factors, organized into four categories (Common Technological Platform, Projects and Applications, Community Interaction, and Expectations and Value of Contribution), provides comprehensive insights into critical concerns affecting DX in SECO. This knowledge base serves as an empirical foundation to guide both researchers and practitioners in addressing DX-related challenges in fostering developer adoption and sustained participation.

These insights were instrumental in shaping our conceptual model. By understanding which factors most strongly influence developer engagement and retention, we incorporated them as key constructs, ensuring that the model reflects not only transparency aspects but also the role of DX in mediating the effectiveness and adoption of transparent practices. Further details on this SMS and Delphi study are available in Zacarias et al.~\cite{zacarias2025Dxfactors}.

\subsubsection{Step 1.2: Analyzing a previous study on transparency in SECO}
We analyzed results from our previous SMS and field study to characterize conditioning factors for transparency in SECO. The SMS was designed to map and explore the state-of-the-art on transparency in SECO, addressing the RQ: ``\textit{How is transparency in the SECO context characterized?}''. We searched major scientific databases and selected 23 primary studies. The findings provided a broad view of transparency-related solutions, conditioning factors, common SECO procedures, and concerns in SECO, particularly in areas related to information access, communication channels, and requirements engineering.

To further explore the practical implications of these factors, we conducted a field study investigating their relevance from the perspective of software developers. This work aimed to answer the question: ``\textit{What is the importance level of conditioning factors for transparency according to developers in the context of an OSSECO?}''. Sixteen software developers who use GitHub, an open platform widely adopted by OSSECO~\cite{LunguLanza2010}, participated in interviews. The results confirmed the applicability and perceived importance of the factors identified in the SMS, as we present in Section~\ref{sec:secoTransDXBack}, and also revealed benefits developers associate with transparent practices. 

These two studies served as the foundation for defining transparency in SECO within our work. Their combined results informed the selection and structuring of the key transparency constructs in our conceptual model. By synthesizing the conditioning factors, common SECO procedures, and perceived benefits of transparency, we established the initial theoretical basis for the model proposed in this article. Further details on our SMS and field study are available in Zacarias et al.~\cite{Zacarias2024}.

\subsubsection{Step 1.3: Analyzing SECO meta-model}
To complement the empirical results from our previous studies on transparency and DX in SECO, we analyzed the SECO meta-model proposed by Wouters et al.~\cite{Wouters2019}. This meta-model provides a comprehensive structure describing the core entities and relationships within SECO and establishes a shared vocabulary for researchers. It is organized into five themes: \textit{actors and roles}, \textit{products and platforms}, \textit{boundaries}, \textit{ecosystem health}, and \textit{strategy}. We focused on the two themes most relevant to our research objectives: \textit{actors and roles} and \textit{products and platforms}, as they align directly with the results of our studies on transparency and DX.

From the \textit{actors and roles} theme, we selected the entities \textit{actor}, \textit{role}, \textit{keystone}, \textit{hub}, and \textit{niche player} to represent SECO stakeholders, their responsibilities, and interactions. These concepts are central to understanding both transparency practices and DX dynamics. For instance, transparency factors such as communication channels and information reliability often depend on the keystone’s role, while DX factors are shaped by interactions between niche players and the platform.

From the \textit{products and platforms} theme, we included the entities \textit{product} and \textit{platform}, which represent the core technological environment where transparency mechanisms and DX are manifested. Aspects such as transparent documentation, tool access, and usability, highlighted in our prior findings, are closely tied to these entities. The remaining themes (\textit{boundaries}, \textit{ecosystem health}, and \textit{strategy}) were excluded as they fell outside the scope of this work, which centers on modeling the interplay between transparency and DX in SECO, although they may be relevant in future extensions.

By combining the relevant elements of the SECO meta-model with empirical insights from our SMS and field studies, we established a coherent conceptual foundation for defining the constructs and relationships of our model. This integration ensures theoretical robustness while preserving practical relevance to transparency and DX in SECO.

\subsection{Step 2: SECO-TransDX design}\label{sec:design}
In this step, we specified the proposed solution to address the research problem~\cite{Wieringa2014}. We adopted the theory-building framework proposed by Sjøberg et al.~\cite{Sjøberg2008}, which organizes this process into three sub-steps: (i) Step 2.1: Defining constructs, which identifies the key elements that constitute the theory; (ii) Step 2.2: Defining propositions, which describes the relationships established among these constructs; (iii) Step 2.3: Providing explanation, which contextualizes the propositions and presents the evidence supporting them; and (iv) Step 2.4: Defining scope, which specifies the boundaries within which the theory is applicable.

\begin{figure}[!ht]
\includegraphics[width=0.7\linewidth]{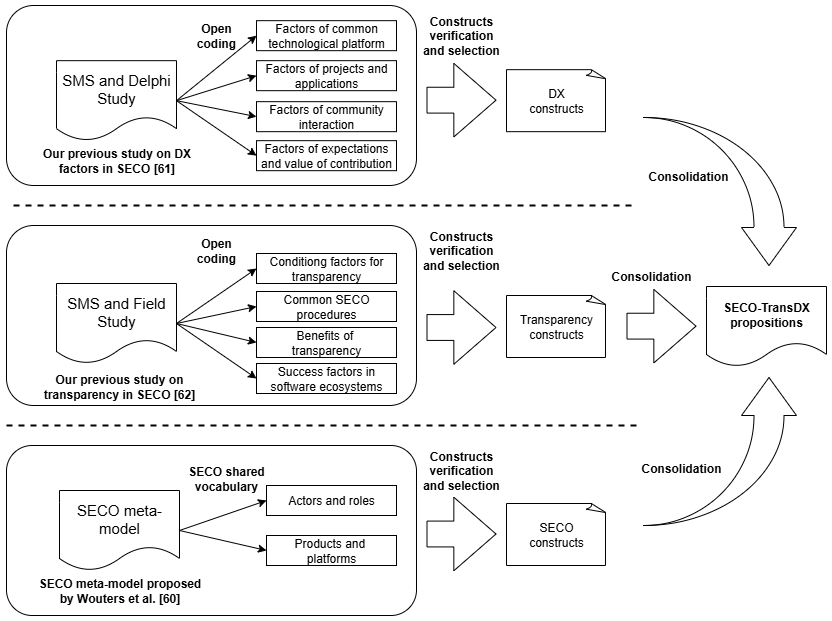}
\caption{The terms identification and selection.}
\label{fig:terms_identification_and_selection}
\Description{Terms identification and selection.}
\end{figure}

\paragraph{Step 2.1: Defining constructs.}
We identified the constructs that would form the foundation of the SECO-TransDX model. These constructs were derived from the results of our previous studies on DX and transparency in SECO~\cite{zacarias2025Dxfactors, Zacarias2024}, as well as from the SECO meta-model proposed by Wouters et al.~\cite{Wouters2019}. Fig.~\ref{fig:terms_identification_and_selection} illustrates the flow from construct identification to consolidation.

From our previous study on DX in SECO~\cite{zacarias2025Dxfactors}, open coding yielded four main groups of factors: ``Common Technological Platform'', ``Projects and Applications'', ``Community Interaction'', and ``Expectations and Value of Contribution''. These were then verified and selected as DX constructs for the conceptual model. Similarly, our study on transparency in SECO~\cite{Zacarias2024} generated, through open coding, a set of conditioning factors for transparency, common SECO procedures, benefits of transparency, and success factors in SECO. After verification, these elements were consolidated as transparency constructs.

The SECO meta-model provided a shared vocabulary and structural reference to ensure conceptual alignment with the broader SECO literature. We examined its five thematic areas and selected the entities ``actors and roles'' and ``products and platforms'' as SECO constructs, since they directly relate to the context in which transparency and DX emerge and interact.

During the verification and selection step, the first author identified the candidate codes to be treated as constructs, and the other two authors checked and reviewed these candidates to ensure relevance, clarity, and consistency. For example, in our previous study on transparency~\cite{Zacarias2024}, the open coding process identified the code ``The existence of communication channels between actors and keystone''. This was validated as relevant by all authors and formalized as the construct \textit{Existence of Communication Channels}. This construct was then incorporated into the SECO-TransDX model as part of the transparency dimension, where it relates to DX constructs such as \textit{Community Interaction}.

In this step, we initially identified 56 concepts from previous studies and the SECO meta-model that were considered potential constructs related to DX, transparency, and SECO. These constructs were then verified, refined, and consolidated to define the propositions of SECO-TransDX, ensuring both theoretical robustness and empirical grounding. The results of this step are presented in Section~\ref{sec:model}.

\paragraph{Step 2.2: Defining propositions.}
We defined the propositions of SECO-TransDX, which represent the relationships among its constructs~\cite{Sjøberg2008}. Propositions explain phenomena by associating constructs and describing how they interact within the model~\cite{Sjøberg2008}, being a core element of conceptual models~\cite{MALCHER2025107672}. In our case, the SECO-TransDX propositions capture specific fragments of the conceptual model and emerged from empirical evidence gathered in our previous SMS, field study, and Delphi study. These propositions provide a structured means of linking transparency, DX, and SECO constructs, thereby enabling a coherent representation of how transparency practices influence DX within SECO.

\paragraph{Step 2.3: Providing explanation.}
We provided evidence-based explanations from the literature and our empirical studies for each SECO-TransDX proposition. These explanations contextualize the relationships among constructs and justify their inclusion in the model. For instance, for proposition P3 (\textit{Third-party developers interact with a software ecosystem portal to consume information about a common technological platform}), we provided the following explanation: ``\textit{SECO portals are the main source of information related to the artifacts that constitute a common technological platform~\cite{meireles2019instrumento}. Third-party developers interact with SECO portals to consume this information and to be aware of the processes and elements (e.g., documentation, source code, repositories, tools etc.) that are part of the common technological platform. Hence, they are able to develop software components and make their contributions to a SECO~\cite{Souza2020, Parracho2024}}''. Such evidence came from the primary studies in our SMS, the results of our field study, and the Delphi evaluation. The detailed explanations for each proposition are presented in Section~\ref{sec:model}.

\paragraph{Step 2.4: Determining scope.}
We determined the scope of the SECO-TransDX conceptual model by defining the boundaries within which the model is applicable. We assume that SECO-TransDX is suitable for different types of SECO (PSECO, OSSECO, and hybrid) since our previous studies on transparency and DX, as well as the SECO meta-model analysis, did not target a specific SECO type. This broad applicability suggests that the model’s constructs and propositions may be relevant across different governance models, technological architectures, and participation structures.

\subsection{Step 3: SECO-TransDX evaluation}\label{sec:evaluation}
This step aimed to assess whether the proposed solution—in our case, SECO-TransDX—achieves the intended objectives. For this purpose, we employed the Delphi study~\cite{DALKEY1969408}, a structured and iterative approach designed to obtain consensus from a panel of experts. The Delphi study involves multiple rounds of data collection and analysis, in which participants remain anonymous and can revise their responses based on group feedback, gradually converging toward agreement~\cite{WANG2022100463, MALCHER2025107672}. This method is particularly suited for evaluating solutions in emerging and complex domains such as SECO, where empirical evidence may be limited, as it enables the systematic validation of ideas and propositions using expert judgment~\cite{Alarabiat2019, OLIVERO2022106874}. In our study, we adopted the process outlined by Olivero et al.~\cite{OLIVERO2022106874} and Malcher et al.~\cite{MALCHER2025107672}, as detailed in the following subsections.

\subsubsection{Step 3.1: Planning evaluation}\label{sec:planning}
This step is divided into the following activities: selecting
subject, selecting experts, and statistical processing~\cite{OLIVERO2022106874, MALCHER2025107672}. 

\paragraph{Selecting subject.}
The quality of the Delphi study results is influenced by selecting an appropriate subject and designing the questionnaire effectively~\cite{OLIVERO2022106874}. This step involved defining the RQ that guided the questionnaire’s purpose and structure, as well as establishing the criteria for expert selection. In our case, we defined two research questions to evaluate SECO-TransDX:

\begin{itemize}
    \item Delphi-RQ1: Do the experts agree with the propositions presented for SECO-TransDX?
    \item Delphi-RQ2: Do the experts agree that SECO-TransDX meets the criteria of ambiguity, explanatory power, parsimony, generality, and utility?
\end{itemize}

Based on these RQ, the main purpose of the questionnaire was to assess experts’ agreement with the propositions of SECO-TransDX and to verify compliance with the evaluation criteria proposed by Sjøberg et al.~\cite{Sjøberg2008}. As noted by Malcher et al.~\cite{MALCHER2025107672}, propositions can serve as the basis for evaluation because they represent the conceptual model as a whole, including its constructs, the relationships among them, and its diagrammatic representation. According to Sjøberg et al.~\cite{Sjøberg2008}, the criteria of ambiguity, explanatory power, parsimony, generality, and utility are particularly relevant for evaluating empirically grounded software engineering theories, such as SECO-TransDX.

The questionnaire was structured into three parts and was preceded by a brief introduction presenting the study’s objectives and the informed consent form, which participants were required to read and agree to before proceeding. Part I contained five demographic questions aimed at outlining the experts’ profiles, enabling us to contextualize their responses according to their background and experience with SECO, software transparency, DX, and software development. Part II comprised eight closed-ended questions addressing the propositions of SECO-TransDX, each accompanied by an optional open-ended question to capture additional comments or suggestions. This part was designed to assess the degree of expert agreement with the model’s propositions, which collectively represent its constructs, relationships, and overall structure (as detailed in Section~\ref{sec:model}). Part III included five closed-ended questions related to the overall evaluation criteria for conceptual models as defined by Sjøberg et al.~\cite{Sjøberg2008}, together with five optional open-ended questions to allow participants to elaborate on their assessments. This final part aimed to evaluate the model’s theoretical soundness, practical applicability, and relevance across different SECO contexts.

We conducted a pilot to evaluate the questionnaire with a professional with 12 years of experience in project management and an active research background in SECO. The aim was to verify the clarity, coherence, and effectiveness of the questions before distributing them to the expert panel. The pilot participant took approximately 30 minutes to complete the questionnaire, and the feedback led to minor refinements to improve its suitability. These refinements included adjusting layout spacing, clarifying the description of knowledge levels (including level 0 for consistency), improving the readability of a table, adding contextual definitions for study areas to assist participants unfamiliar with SECO, ensuring scale definitions were unambiguous, verifying section titles, and standardizing the use of acronyms such as SECO and DX. After carrying out the pilot, we made the necessary adjustments, and the final version of the questionnaire is available in the supplementary material at \url{https://doi.org/10.5281/zenodo.16898933}.

\begin{figure}[!ht]
\includegraphics[width=0.5\linewidth]{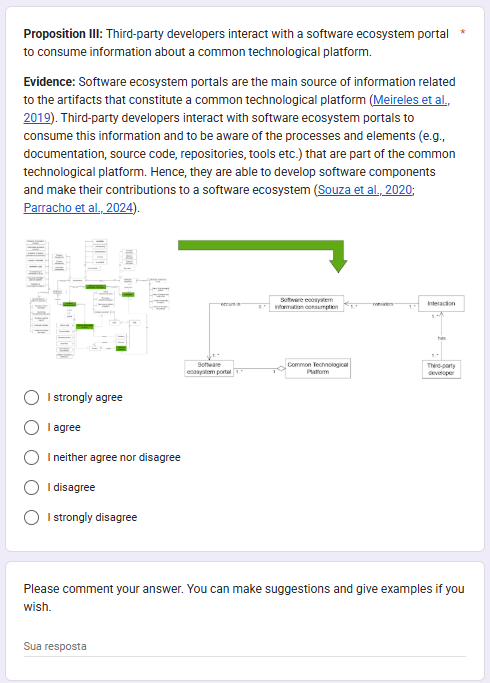}
\caption{Example of a question that was measured using a 5-point Likert scale.}
\label{fig:example-p3}
\Description{Example of a question that was measured using a 5-point Likert scale.}
\end{figure}

\begin{figure}[!ht]
\includegraphics[width=0.5\linewidth]{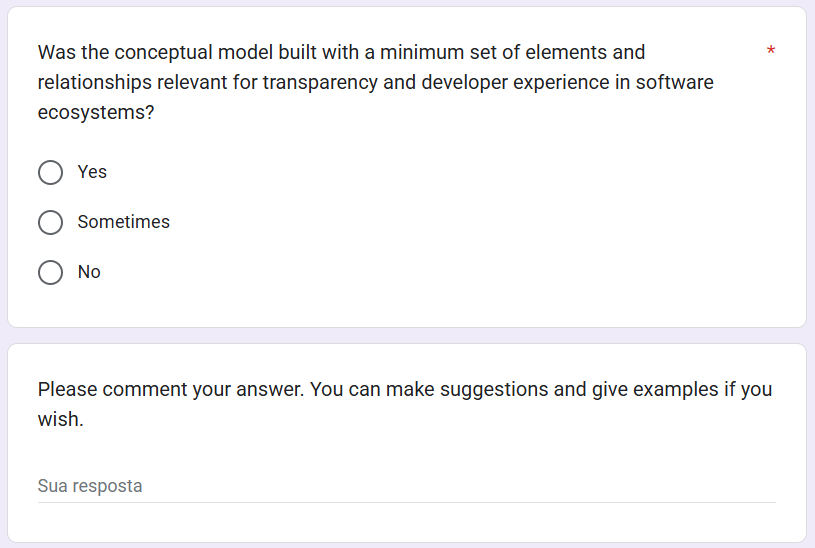}
\caption{Example of a question that was measured using a 3-point Likert scale.}
\label{fig:example-criteria}
\Description{Example of a question that was measured using a 3-point Likert scale.}
\end{figure}

The questionnaire was made available to the expert panel via a Google Forms\footnote{https://docs.google.com/forms/} link. To preserve anonymity, the composition of the panel was not disclosed, and all communications were conducted individually. Responses were anonymized in the reports for each round by removing any identifying information about the participants. Fig.~\ref{fig:example-p3} shows a 5-point Likert scale question used to evaluate the propositions of the SECO-TransDX model, highlighting the direct link between each proposition and the model’s conceptual structure. Fig.~\ref{fig:example-criteria} presents an example of a 3-point Likert scale question used to assess the model’s overall evaluation criteria. For instance, P3 aligns directly with the concepts ``Common technological platform'', ``Software ecosystem portal'', ``Software ecosystem information consumption'', ``Interaction'', and ``Third-party developer'' and their relationships.

\paragraph{Selecting experts.}
The careful selection of experts is a critical factor for the success of the Delphi study, as it directly affects the quality and reliability of the results~\cite{Alarabiat2019, Torrecilla-Salinas2019}. Nonetheless, defining an appropriate expert profile is one of the main challenges in Delphi studies~\cite{Alarabiat2019}. The literature offers no universal criteria for establishing the ideal participant profile, although aspects such as professional experience, subject-matter expertise, and willingness to engage throughout the process are frequently considered relevant~\cite{Alarabiat2019, Torrecilla-Salinas2019}. Similarly, there is no agreement on the optimal panel size. Previous studies report that Delphi panels often include between 15 and 20 experts to answer a structured questionnaire~\cite{OLIVERO2022106874, MALCHER2025107672}. As suggested by Torrecilla-Salinas et al.~\cite{Torrecilla-Salinas2019}, the panel size should be large enough to ensure representativeness while remaining small enough to keep the process efficient and manageable.

As our goal was to evaluate the SECO-TransDX conceptual model, the ideal expert panel would include professionals with expertise in both SECO and DX, as well as familiarity with transparency-related practices and software development. However, identifying specialists with deep knowledge in all these areas proved challenging, given that the SECO research community is relatively small compared to more mature fields in software engineering~\cite{Mens2023, Santos2023}. To address this, we adopted an approach similar to that of Torrecilla-Salinas et al.~\cite{Torrecilla-Salinas2019}, selecting participants with some experience in at least one of the domains relevant to the evaluation, complemented by theoretical knowledge of the others. This strategy allowed us to form a heterogeneous panel, reflecting the diversity of backgrounds and perspectives typically found in the SECO community.

We invited potential participants from the active communities of SECO, DX, and software transparency research, as well as members of program committees and organizers of related events, such as the International Workshop on Software Engineering for Systems-of-Systems and Software Ecosystems (SESoS) and the International Conference on Cooperative and Human Aspects of Software Engineering (CHASE). In addition, we applied convenience sampling~\cite{Baltes2022} to reach professionals from our network with experience in SECO development or management activities, contacting them via email and professional communication channels (e.g., WhatsApp and LinkedIn). We initially invited 50 experts to participate, receiving 30 confirmations from individuals who completed the questionnaires throughout the Delphi rounds. 

In line with Alarabiat and Ramos~\cite{Alarabiat2019}, we requested participants to self-assess their expertise in four knowledge areas relevant to the evaluation of SECO-TransDX: SECO, software transparency, DX, and software development. The assessment was conducted through a demographic questionnaire that included a six-point scale ranging from 0 (none) to 5 (very high). The scale was anchored with descriptors to guide consistent responses: 0 = no experience, 1 = studied in class or in books (very low), 2 = used in classroom projects (low), 3 = applied in own projects (average), 4 = used in few industry projects (high), and 5 = used in several industry projects (very high). In addition, we collected information on participants’ academic qualifications (secondary or technical school, bachelor’s, master’s, specialization, Ph.D.), sector (public, private, both, or independent), and professional career (academia, industry, or both). Table~\ref{tab:DelphiDemographic} presents a summary of information characterizing the profile of the participants. We assigned an identifier (ID) to each expert, following the order of the interviews carried out (E1 to E30), to identify them throughout this article.

\begin{table}[!ht]
\footnotesize
\caption{Characterization of the experts.}
\label{tab:DelphiDemographic}  
\begin{tabular}{llllllll}
\hline
\multirow{2}{*}{ID} & \multirow{2}{*}{Academic Qualification} & \multirow{2}{*}{Sector} & \multirow{2}{*}{Professional Career} & \multicolumn{4}{c}{Experience Level}                                                                                                                                                                                                                               \\ \cline{5-8} 
                    &                                         &                         &                                      & \begin{tabular}[c]{@{}l@{}}Software\\ Ecosystems\end{tabular} & \begin{tabular}[c]{@{}l@{}}Software \\ Transparency\end{tabular} & \begin{tabular}[c]{@{}l@{}}Developer\\ Experience\end{tabular} & \begin{tabular}[c]{@{}l@{}}Software\\ Development\end{tabular} \\ \hline
                E1  &   Ph.D.                                  &     Public             &  Academia              &
        Very Low    &  Low                                     &  Average               &  High                                   \\
             E2  &  Bachelor's degree                          &   Private               &   Academia and Industry                  &
        Average    &       Average                            &    Average            &      High                              \\ 
             E3  &   Master's degree                         &   Public               &  Academia and Industry             &
        Low    &    Very Low                                &  Very Low               &   Very Low                               \\ 
             E4  &   Bachelor's degree                      &  Independent developer   &  Academia            &
        Low    &    Very Low                                   &  Low              &     Average                               \\ 
             E5  &    Ph.D.                                 &   Public               &    Academia            &
        Low    &    Average                                   &  Average              &     Average                               \\ 
             E6  &   Master's degree                                   &   Private               &  Industry              &
        High    &   Average                                    &   High             &     Very High                               \\ 
             E7  &     Master's degree                           &    Private              &   Industry            &
        Very High    &   Very High                                    &   High             &  High                                  \\ 
             E8  &    Master's degree                                 &  Private                &   Academia and Industry              &
        High    &     High                                  &    High            &    Very High                             \\ 
             E9  &   Ph.D.                                 &   Public               &    Academia            &
       Average     &  Low                                     &    High            &    Very High                                \\ 
             E10  &    Ph.D.                                 &   Public               &    Academia                &
       Low     &      Low                                 &    High            &   Very High                                 \\ 
             E11  &    Master's degree                     &   Public               &   Academia            &
       None    &    None                                   &    Low            &     Very Low                               \\ 
             E12  &    Ph.D.                                 &   Public               &    Academia         &
        Low    &    Low                                   &   Low             &   Very High                                 \\ 
             E13  &  Ph.D.                                 &   Public               &    Academia        &
       Very High     &       High                                &    Very High            &     Very High                               \\ 
             E14  &    Specialization degree                 & Public              &   Industry            &
       Low     &     Very Low                                  &     High           &   Very High                                 \\ 
             E15  &    Ph.D.                                 &     Private             &     Industry          &
       Very Low     &    Low                                   &  Very High               &    Very High                                 \\ 
             E16  &    Ph.D.                                 &   Public               &    Academia          &
        Low    &        None                               &    Very High            &     Very High                               \\ 
             E17  &    Bachelor's degree                  &   Public               &  Academia and Industry             &
        High    &       Very Low                                &   Low             &     Very High                               \\ 
             E18  &      Ph.D.                                 &   Public               &    Academia          &
      Very High      &       Average                           &   Low             &       Average                             \\ 
             E19  &      Ph.D.                                 &   Public               &    Academia      &
      Low      &         Very High                              &      Very High            &      Very High                                \\ 
             E20  &    Ph.D.                                 &   Public               &    Academia and Industry      &
      Very Low      &       Low                                &     Low           &      Very High                               \\ 
             E21  &     Bachelor's degree                     &  Private                &   Industry            &
      Low      &    Average                                   &    Low            &    Average                                \\ 
             E22  &       Bachelor's degree                      &   Public               &   Academia and Industry            &
      Average      &     Very High                                  &     Average           &         Very High                           \\ 
             E23  &  Ph.D.                                 &   Public               &    Academia         &
       High     &         Average                              &     Average           &      High                              \\ 
             E24  &     Master's degree                        &    Public               &    Academia          &
     Average       &          High                             &       High         &         High                           \\ 
             E25  &       Master's degree                      &    Public              &    Academia and Industry           &
     High       &         Average                              &     Very High            &         Very High                            \\ 
             E26  &      Ph.D.                                 &   Public               &    Academia          &
      Very Low       &       Very Low                                 &    Very Low             &     Low                                \\ 
             E27  &     Bachelor's degree                      &    Public and Private              &  Industry             &
     Average       &        Low                               &    Low            &     High                               \\ 
             E28  &    Bachelor's degree                                 &    Private              &    Industry           &
     Very High       &     Very High                                  &     Very High           &     Very High                               \\ 
             E29  &     Ph.D.                                 &   Public               &    Academia   &
      Very High        &         Low                              &     Average           &      High                              \\ 
             E30  &     Ph.D.                                 &   Public               &    Academia        &
    Average        &    Very Low                                  &      Very Low          &       Very High                             \\  \hline
\end{tabular}
\end{table}

The expert panel comprised 30 participants with diverse profiles. In terms of academic qualifications, 37\% held a Ph.D., 27\% a Master’s degree, 33\% a Bachelor’s degree, and 3\% a Specialization degree. Most participants worked in the public sector (67\%), followed by the private sector (23\%) and mixed public/private experience (10\%). Professionally, 60\% were from academia, 23\% from industry, and 17\% from both.

Expertise levels varied across domains. In SECO, 40\% reported high or very high experience, while 33\% had average experience and 27\% had low or very low. For software transparency, 33\% had high or very high expertise, 30\% average, and 37\% low or very low. DX showed a similar pattern, with 47\% reporting high or very high expertise, 23\% average, and 30\% low or very low. In software development, the highest proficiency was observed, with 73\% at high or very high levels, 17\% average, and only 10\% low or very low. This balanced distribution of expertise reflects the heterogeneous nature of the SECO community and ensures a broad, complementary basis for evaluating the model.

\paragraph{Statistical processing.}
In Delphi studies, statistical processing involves defining in advance the methods to analyze experts’ responses in each round, as well as the stopping criteria that determine when additional rounds are no longer required. Establishing these procedures beforehand helps to avoid potential bias in interpreting the results~\cite{OLIVERO2022106874}. The analysis of responses across rounds seeks to determine whether consensus has been achieved. Several statistical techniques can be used in Delphi research to measure consensus~\cite{VONDERGRACHT20121525, Chalmers2019}. In our study, we employed descriptive statistical analysis to assess the level of agreement among experts.

Descriptive analysis was applied to examine the frequency distribution and central tendency of the responses~\cite{thompson_descriptive_2009}. The Delphi questionnaire employed both a 5-point and a 3-point Likert scale, with responses converted into numerical values for analysis. The 5-point scale was used for items requiring a more fine-grained understanding of the panelists’ opinions (propositions), while the 3-point scale was adopted for items assessing the overall evaluation criteria, providing a simpler and clearer choice framework, reducing response fatigue, and maintaining focus.

For each item, we calculated the median, mode, standard deviation (SD), and interquartile range (IQR), measures widely used to assess consensus in Delphi studies~\cite{Giannarou2014}. These statistics supported the identification of consensus levels within the expert panel, as well as the degree of agreement on each proposition and on the overall evaluation criteria.

The IQR, representing the distance between the 25th and 75th percentiles, was adopted due to its robustness as a consensus indicator~\cite{Alarabiat2019, MALCHER2025107672}. Smaller IQR values indicate higher agreement among experts~\cite{Giannarou2014}. The median was used to determine the strength of panel support for each item, with higher values reflecting stronger agreement~\cite{Giannarou2014}. The SD was analyzed to observe the dispersion of responses, complementing the IQR, with values below 1.5 considered evidence that consensus had been achieved~\cite{Giannarou2014, MALCHER2025107672}.

Stopping conditions in Delphi studies can be established based on consensus, stability, or a predefined number of rounds~\cite{OLIVERO2022106874}. In our study, we adopted consensus as the stopping condition, defined by both the IQR and SD, following approaches used in previous Delphi research~\cite{Alarabiat2019}. Specifically, consensus was considered achieved when $IQR \leq 1$ and $SD \leq 1.5$. Additionally, agreement or disagreement on each proposition or overall evaluation criterion of the conceptual model was assessed using the strategy proposed by Olivero et al.~\cite{OLIVERO2022106874}, in which experts are considered to agree with a given item if more than 51\% assign it a score of 4 or 5 on the Likert scale. This combined approach allowed us to evaluate not only the level of consensus but also the direction of agreement among the panelists.

\subsubsection{Step 3.2: Conducting and reporting evaluation}\label{sec:conducting}
This step encompassed three main activities: first round, generating statistical data, and Further rounds~\cite{OLIVERO2022106874, MALCHER2025107672}.

\paragraph{First round.} 
This stage introduced the finalized questionnaire to the expert panel. Following the approach recommended by Olivero et al.~\cite{OLIVERO2022106874} and used by Malcher et al.~\cite{MALCHER2025107672}, we opted for the use of a well-structured questionnaire rather than a preliminary version for a test round, as its design had already been refined through a prior pilot with a practitioner with 12 years of experience in project management and an active research background in SECO. This pilot ensured the clarity, coherence, and suitability of the instrument before distribution.

\paragraph{Generating statistical data.}
Responses from the experts were analyzed using the statistical procedures defined in advance (Section~\ref{sec:planning}). An anonymized summary of the panel’s feedback was prepared after each round, detailing the central statistical measures and consolidating any comments provided. This process was essential for determining whether the pre-established stopping conditions had been met~\cite{OLIVERO2022106874}.

\paragraph{Further rounds.}
Additional rounds were conducted only if the stopping conditions were not reached in the preceding stage. In such cases, participants received the anonymized summary from the previous round, which included the aggregated statistical results and selected comments from other experts. This allowed them to review the group’s overall position and either maintain or revise their initial responses. In line with Delphi study best practices~\cite{Torrecilla-Salinas2019, OLIVERO2022106874}, two to three rounds are typically sufficient to achieve consensus within expert panels.

\subsubsection{Step 3.3: Determining results and conclusions}\label{sec:conducting}
Once the predefined stopping conditions were satisfied, we concluded the Delphi process and produced a final report. This report consolidated the statistical analyses and qualitative feedback obtained throughout the rounds, providing a comprehensive interpretation of the expert panel’s evaluations. The results served both to validate the SECO-TransDX constructs and propositions and to inform refinements in the conceptual model, ensuring its robustness and applicability to the SECO context.

\section{SECO-TransDX conceptual model}\label{sec:model}
\begin{figure}[!ht]
\includegraphics[width=1.0\linewidth]{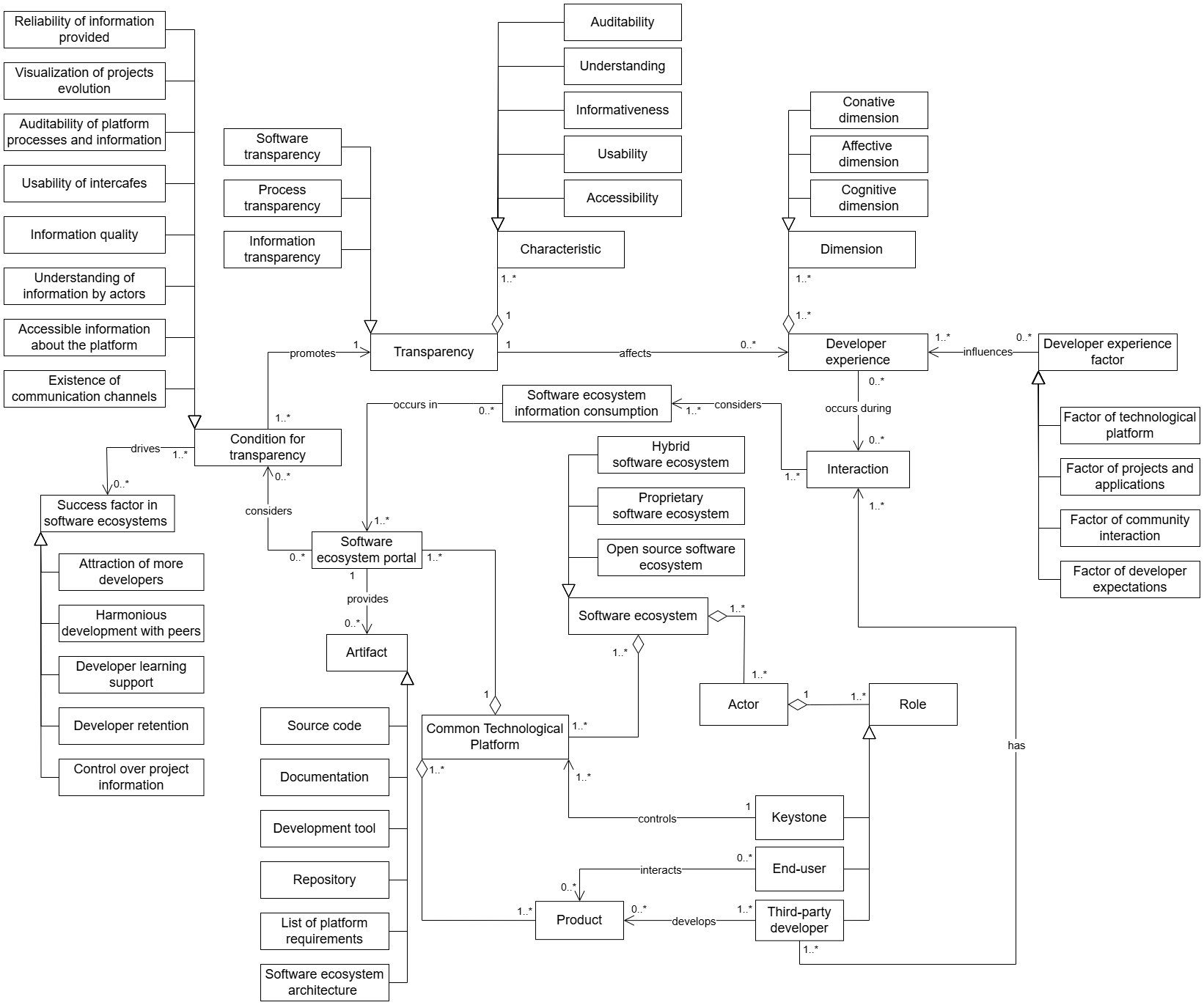}
\caption{First version of the SECO-TransDX conceptual model: \textit{Transparency in Software Ecosystems from a Developer Experience Perspective}.}
\label{fig:SECO-TransDX_v1} 
\Description{A conceptual model design with UML notation.}
\end{figure}

This section presents the first version of the SECO-TransDX conceptual model, which is composed of a diagrammatic representation, a glossary of constructs, and a set of propositions. Fig.~\ref{fig:SECO-TransDX_v1} illustrates the SECO-TransDX using a UML class diagram, comprising 56 constructs related to transparency, DX, and SECO, as well as the relationships among them. Table~\ref{tab:propositions_v1} details the propositions that connect these constructs. The glossary describing the 56 constructs is provided in the supplementary material at \url{https://doi.org/10.5281/zenodo.16898933}, due to space constraints in the article.

\begin{table}[!ht]
\footnotesize
\caption{Initial version of the statement of SECO-TransDX propositions}
\label{tab:propositions_v1}      
\begin{tabular}{p{0.03cm} p{6cm} p{8cm}}
\hline\noalign{\smallskip}
ID & Proposition & Explanation  \\
\noalign{\smallskip}\hline\noalign{\smallskip}

P1 & \underline{Software ecosystem portals} are web interfaces that provide \underline{artifacts} from a \underline{common technological platform}, considering the type of \underline{software ecosystem}. & A SECO can be defined as a set of actors that interact with a common technological platform, which allows the exchange of information, resources, and artifacts~\cite{Jansen2009, Wouters2019}. In such a context, SECO portals are web interfaces that provide artifacts from a common technological platform, such as source code, documentation, development tools, repositories, a list of platform requirements, and SECO architecture~\cite{meireles2019instrumento}. The artifacts available on the portals vary according to the type of SECO, which can be classified as open source, proprietary, or hybrid~\cite{Manikas2016}. \\ \hline

P2 & \underline{Third-party developers} are one of the \underline{roles} of \underline{actors} in \underline{software ecosystems} who interact with a \underline{common technological platform}, which is controlled by a \underline{keystone}, to develop their own \underline{products} for \underline{end-users}. & SECO are based on the collaboration of multiple actors~\cite{Manikas2016}. We can identify three main actors’ roles: keystone, end-users, and third-party developers~\cite{Hanssen2012}. A keystone is an organization or group that drives the development of a common technological platform~\cite{Lewellen2020}. End-users are customers who adopt or intend to adopt a software product to make them more productive~\cite{hou_systematic_2022}. Finally, third-party developers use the platform as a basis to develop new products and solutions~\cite{choi_where_2020}. \\ \hline

P3 & \underline{Third-party developers} interact with a \underline{software ecosystem portal} to \underline{consume information} about a \underline{common technological platform}. & SECO portals are the main source of information related to the artifacts that constitute a common technological platform~\cite{meireles2019instrumento}. Third-party developers interact with SECO portals to consume this information and to be aware of the processes and elements (e.g., documentation, source code, repositories, tools etc.) that are part of the common technological platform. Hence, they are able to develop software components and make their contributions to a SECO~\cite{Souza2020, Parracho2024}. \\ \hline

P4 & \underline{Software ecosystem portals} consider \underline{conditions} to promote transparency in \underline{software ecosystems}. & In SECO, transparency is considered from the software perspective, i.e., a condition that encompasses openness, clarity, and visibility of the mechanisms, processes, and actions of software applications~\cite{LeiteCappelli2010, Isong2022}. To do so, SECO portals consider a set of conditions to promote transparency: existence of communication channels, accessible information about the platform, understanding of information by actors, information quality, usability of interfaces, auditability of platform processes and information, visualization of projects evolution, and reliability of the information provided~\cite{Linåker2016, Souza2020, RUNESON2021111088, BEELEN2022102733}. \\ \hline

P5 & \underline{Conditions for transparency} in the \underline{portals} can drive \underline{success factors in software ecosystems}. & When SECO portals can achieve the conditions for transparency, they can drive success factors, i.e. positive advantages of promoting transparency in SECO~\cite{Cataldo2010, Zacarias2024}. Some examples of success factors are attracting more developers~\citep{BEELEN2022102733}, harmonious development with peers~\citep{Zacarias2024}, supporting developer learning of market mechanisms~\cite{Dabbish2012}, developer retention~\cite{meireles2019instrumento}, and control over project information~\cite{Linåker2016}. The success factors contribute to the sustainability of SECO~\cite{fontao2021developer}. \\ \hline

P6 & \underline{Transparency} is supported by a set of \underline{characteristics} related to \underline{accessibility}, \underline{usability}, \underline{informativeness}, \underline{understandability}, and \underline{auditability}, and can be analyzed in \underline{software ecosystems} from \underline{information}, \underline{process}, and \underline{software perspectives}. & Considering related needs in the software context, which includes SECO, Leite and Cappelli~\cite{LeiteCappelli2010} define transparency as the union of characteristics that contribute to its formation and the open information flow: accessibility, usability, informativeness, understandability, and auditability. Furthermore, as a non-functional requirement, transparency should be considered at all stages of software design to clarify how software provides transparency in organizational processes and information~\citep{Cysneiros2013, Hosseini2015}. Hence, transparency can be analyzed in SECO from information, process, and software perspectives~\citep{Parracho2024}. \\ \hline

P7 & \underline{Developer experience} is comprised of \underline{dimensions} and considers a set of \underline{factors} in \underline{software ecosystems}. &  According to Fagerholm and Münch~\cite{Fagerholm2012}, DX is composed of the following dimensions: cognitive, affective, and conative. In SECO, some specific factors influence DX. These factors can be categorized based on the dimensions of DX and characteristics related to the dynamics of SECO: (a) factors of technological platform, which are related to the technical infrastructure for development provided by a common technological platform~\cite{Kauschinger2021, STEGLICH2023111808}, (b) factors of projects and applications, which are related to the process of developing and distributing applications on a common technological platform~\cite{Choia2017Impacts, STEGLICH2023111808}, (c) factors of community interaction, which are related to the interaction between a developer and other developers who integrate a SECO community~\cite{Fontao2020, STEGLICH2023111808}, and (d) factors of developer expectations, which are related to expectations and benefits obtained by a developer’s contribution in interacting with a SECO~\cite{DeSouza2016, STEGLICH2023111808}.  \\ \hline

P8 &  \underline{Transparency} affects the \underline{developer experience} of \underline{third-party developers} during their \underline{interactions} to \underline{consume information} in \underline{software ecosystem portals}. & Third-party developers access SECO portals to consume information about how to develop applications in this ecosystem~\cite{meireles2019instrumento, Parracho2024}. If this information cannot be accessed or understood easily, these third-party developers may have difficulty developing their applications autonomously. This situation is related to the level of transparency in these portals~\cite{Parracho2024}. Lack of transparency contributes to increased barriers to entry for a SECO and a cumbersome start for newcomer third-party developers. This issue can negatively affect the DX and cause software developers to give up interacting with the common technological platform~\cite{SantosEtAl2016, knauss2018continuous, meireles2019instrumento, fontao2021developer}. \\
\noalign{\smallskip}\hline
\end{tabular}
\end{table}

A conceptual model aims to express the meaning of concepts used by domain experts to discuss the problem, establish the relationships between them, communicate effectively, and abstract the domain’s complexity. It also helps clarify ambiguous terms, reducing misunderstandings that may arise from different interpretations of terminology~\cite{wen_modeling_2012, OLIVEIRAJR2020301014}.

Static/structural conceptual models, such as SECO-TransDX, represent domain entities and their relationships. According to Oliveira et al.~\cite{Oliveira2020}, conceptual models for SECO provide a holistic view of the elements, interactions, and governance structures that define an ecosystem’s functioning. Such models capture technical, transactional, and social dimensions, enabling stakeholders to understand the ecosystem’s structure, identify relationships among its components, and support coordinated decision-making. Therefore, SECO-TransDX is designed to represent the relationships and interactions between transparency and DX concepts in SECO, organizing information without prescribing mandatory actions or defining fixed quantitative thresholds for each component.

\section{SECO-TransDX evaluation and refinement}\label{sec:delphi}

The SECO-TransDX model was evaluated through a Delphi study involving 30 experts with experience in software transparency, DX, and/or SECO. The evaluation aimed to gather expert opinions on the model’s propositions and overall evaluation criteria. The study was conducted in two rounds to reach consensus among participants. The study began in December 2024 (first round) and ended in April 2025 (second round).

The questionnaire items were measured using both 3-point and 5-point Likert scales, with responses converted into numerical values for statistical analysis (Table~\ref{tab:likerScale}). Following the approach of Olivero et al.~\cite{OLIVERO2022106874}, two keywords were defined to summarize the panel’s collective decision: agreement and disagreement. Agreement was calculated as the percentage of experts who selected ``I strongly agree'', ``I agree'', or ``Yes'' for a given item, while disagreement was calculated as the percentage who selected ``I strongly disagree'', ``I disagree'', or ``No''. The subsequent subsections detail the execution of the Delphi study, presenting the results and refinements made to SECO-TransDX after each round.

\begin{table}[!ht]
\footnotesize
\caption{Likert scale interpretation}
\label{tab:likerScale}      
\begin{tabular}{l l l}
\hline\noalign{\smallskip}
5-points Likert scale & 3-points Likert scales  & Numeric value  \\
\noalign{\smallskip}\hline\noalign{\smallskip}
I strongly agree & Yes & 5 \\
I agree & - & 4 \\
I neither agree nor disagree & Sometimes & 3 \\
I disagree & - & 2 \\
I strongly disagree & No & 1 \\
\noalign{\smallskip}\hline
\end{tabular}
\end{table}

\subsection{First round}\label{sec:firstRound}
The first round of the Delphi study was conducted between December 2024 and January 2025. In this stage, the experts evaluated the eight propositions of the SECO-TransDX model and answered five questions aligned with the evaluation criteria proposed by Sjøberg et al.~\cite{Sjøberg2008} for the overall assessment of the model. The data collected were subjected to descriptive statistical analysis to examine the results for both the propositions and the evaluation criteria. Additionally, we assessed the reliability of the questionnaire and summarized the findings from this round, as detailed in the following sections.

\begin{figure}[!ht]
\includegraphics[width=1.0\linewidth]{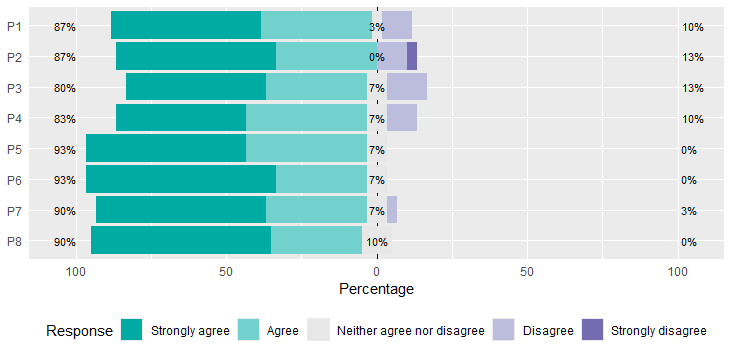}
\caption{Frequency of results for SECO-TransDX propositions evaluation in the first round.}
\label{fig:firstRoundPropositions} 
\Description{Graph.}
\end{figure}

\paragraph{Descriptive analysis of propositions.}
Table~\ref{tab:firstRoundPropositions} summarizes the expert panel’s Round-1 evaluations of the SECO-TransDX propositions, addressing Delphi-RQ1: \textit{``Do the experts agree with the propositions presented for SECO-TransDX?''}. The table reports the median, mode, SD, IQR, and percentages of agreement and disagreement. Strong disagreement was virtually absent, occurring only in P2 (3.3\%). Disagreement levels were comparatively higher for P3 (13.3\%) and P4 (10\%), and to a lesser extent for P1 (10\%), while negligible for P5--P8 (0--3.3\%). Overall agreement was consistently high, with medians $\geq$ 4, modes = 5, IQR = 1 across all propositions, and low SDs ($\approx$0.63--1.10). Fig.~\ref{fig:firstRoundPropositions}  illustrates the response distributions, highlighting the predominance of agreement (\textit{Strongly agree} + \textit{Agree} $\geq$ 80\% for all items) and indicating a ceiling effect toward positive ratings.

\begin{table}[!ht]
\footnotesize
\caption{First-round results for SECO-TransDX propositions.}
\label{tab:firstRoundPropositions}  
\begin{tabular}{lllllllll}
\hline
\multirow{2}{*}{Proposition} & \multirow{2}{*}{Median} & \multirow{2}{*}{Mode} & \multirow{2}{*}{SD} & \multirow{2}{*}{IQR} & \multicolumn{2}{l}{\% Agreement} & \multicolumn{2}{l}{\% Disagreement} \\ \cline{6-9} 
                             &                         &                       &                     &                      & Strongly agree      & Agree      & Disagree     & Strongly disagree    \\ \hline
                        P1     &      4.5                   &   5                    &       0.94              &     1                 &    50\%                 &  36.7\%          &    10\%                   &      0\%       \\
                        P2     &    5.0                      &    5                   &         1.10            &     1                 &      53.3\%                  &     33.3\%       &     10\%                  &     3.3\%        \\
                        P3     &   4.0                        &  5                     &      1.04               &     1                 &      46.7\%               &   33.3\%         &   13.3\%                    &   0\%          \\
                        P4     &    4.0                       &   5                    &       0.95               &    1                  &      43.3\%               &     40\%       &     10\%                  &      0\%       \\
                        P5     &    5.0                      &    5                   &      0.63               &      1                &      53.3\%               &     40\%       &        0\%               &     0\%        \\
                        P6     &     5.0                     &    5                   &       0.63              &     1                 &     63.3\%                &   30\%         &       0\%                &    0\%         \\
                        P7     &   5.0                       &   5                    &        0.77              &    1                  &      56.7\%               &    33.3\%        &       3.3\%                &    0\%         \\
                        P8     &    5.0                      &    5                   &       0.68              &   1                   &     60\%                &    30\%        &      0\%                 &      0\%       \\ \hline
\end{tabular}
\end{table}

To identify the adjustments required for the SECO-TransDX model, we conducted a qualitative analysis of the experts’ open-ended responses to the propositions. First, we applied open coding~\cite{corbin2014basics} to extract relevant excerpts from the answers and assign initial codes. Then, we used axial coding, as described by Corbin and Strauss~\cite{corbin2014basics}, to group the codes into higher-level categories. The first author carried out both coding stages, and the co-authors reviewed the codes and categories to ensure the reliability and validity of the process. The full set of responses, along with the corresponding codes and categories, was compiled in a supplementary document to maintain traceability across multiple iterative review cycles, supported by discussions among the authors and grounded in the literature. Table~\ref{tab:codingExamples} presents illustrative examples of the coding process.

\begin{table}[!ht]
\footnotesize
\caption{Illustration of the coding process.}
\label{tab:codingExamples}      
\begin{tabular}{p{8cm} p{6cm}}
\hline
\multicolumn{2}{p{14cm}}{\textbf{Excerpt on Proposition 1:} ``How open portals facilitate the sharing of APIs and documentation, encouraging collaboration between developers and researchers.'' [E3]} \\ \hline
\textbf{Code} & \textbf{Category} \\ \hline
R1: SECO portals share artifacts that foster collaboration within the ecosystem. & Reinforcement in SECO-TransDX (R) \\ \hline

\multicolumn{2}{p{14cm}}{\textbf{Excerpt on Proposition 2:} ``In practice, they [developers] often develop tools, libraries, or APIs for other developers, for example.'' [E1]} \\ \hline
\textbf{Code} & \textbf{Category} \\ \hline
Ch4: Third-party developers develop and consume artifacts and tools in SECO. & Changes in SECO-TransDX (Ch) \\ \hline

\multicolumn{2}{p{14cm}}{\textbf{Excerpt on Proposition 3:} ``Its success is not guaranteed, as it depends on the portal’s structure, acceptance by the institution or platform holders, and organizational factors.'' [E23]} \\ \hline
\textbf{Code} & \textbf{Category} \\ \hline
NCh6: SECO information consumption depends on the structure of the portal. & No changes in SECO-TransDX (NCh) \\ \hline
\end{tabular}
\end{table}

Building on this analysis, we organized the resulting codes into three groups according to their implications for the conceptual model: reinforcements (R), when the comments supported existing propositions; no changes (NCh), when the comments provided contextual clarifications without requiring adjustments; and changes (Ch), when the comments suggested refinements.

Table~\ref{tab:codesReiforcePropositions} presents the codes reinforcing the propositions. These comments emphasized, for instance, the role of SECO portals as facilitators of collaboration (P1--R1), the recognition of third-party developers as relevant actors in common technological platforms (P2--R2), and the association of transparency with DX and ecosystem success (P4--R5, P6--R7, P8--R9). Taken together, these reinforcements indicate that the SECO-TransDX propositions were generally well aligned with expert perceptions.

\begin{table}[!ht]
\footnotesize
\caption{Codes associated with comments on SECO-TransDX propositions that reinforce (R) propositions in the first round.}
\label{tab:codesReiforcePropositions}   
\begin{tabular}{ccp{12cm}}
\multicolumn{1}{l}{Proposition} & \multicolumn{1}{l}{ID} & Code                                                                                                                       \\ \hline
P1                              & R1                     & SECO portals share artifacts and tools, fostering collaboration within the ecosystem.                                      \\ \hline
P2                              & R2                     & R2: A third-party developer is one of the roles of actors who interact with products of common technological platforms.     \\ \hline
P3                              & R3                     & SECO   portals are the main source of information related to the products of a common technological platform.            \\ \hline
\multirow{2}{*}{P4}             & R4                     & Transparency enables stakeholders to access information and understand procedures in common technological platforms.   \\ \cline{2-3} 
                                & R5                     & Transparency relies on conditioning factors in SECO, which are necessary but not sufficient for its achievement.       \\ \hline
P5                              & R6                     & Transparency in common technological platforms drive success factors in SECO.                                            \\ \hline
P6                              & R7                     & Transparency is supported by accessibility,   usability, informativeness, understandability, and auditability.          \\ \hline
P7                              & R8                     & Some specific factors influence DX based on the characteristics related to the dynamics of the SECO. \\ \hline
P8                              & R9                     & Transparency influences SECO procedures' quality and impacts third-party developers' DX.                                 \\ \hline
\end{tabular}
\end{table}

Table~\ref{tab:codesNoChangesPropositions} summarizes the comments that did not lead to changes in the model. While these observations did not challenge the propositions, they added nuances regarding contextual conditions. For example, experts pointed out that portals are not the sole interface for ecosystem interactions (P1--NCh01), that transparency is shaped by complementary factors such as governance and clarity (P4--NCh07), and that its effects on DX may differ depending on preferences and context (P8--NCh11). Such perspectives nuance the model without altering its structure.

\begin{table}[!ht]
\footnotesize
\caption{Codes associated with comments on SECO-TransDX propositions that do not change (NCh) the conceptual model in the first round.}
\label{tab:codesNoChangesPropositions}   
\begin{tabular}{ccp{7cm}p{5cm}}
\hline
\multicolumn{1}{l}{Criteria} & \multicolumn{1}{l}{ID} & Code                                                                                                                                                          & Justification \\ \hline
\multirow{3}{*}{P1}          & NCh01                  & Portals are not the only interface for interaction in SECO.                                                                                                 &  Portals are one of the main interfaces in SECO~\cite{meireles2019instrumento}             \\ \cline{2-4} 
                             & NCh02                  & Portal design impacts engagement, and disorganization discourages use.                                                                                        & This argument is supported by the evidence presented in  Parracho et al.~\cite{Parracho2024}.             \\ \cline{2-4} 
                             & NCh03                  & Resource completeness and accessibility vary with portal maturity, governance, and audience.                                                              &   SECO-TransDX is limited to improving the
understanding of DX-driven transparency in SECO.            \\ \hline
P2                           & NCh04                  & Internal developers are also actors in SECO.                                                                                                                &  Although SECO-TransDX primarily focuses on third-party developers, it also clarifies that other actors can be part of an ecosystem\cite{Manikas2016, Wouters2019}.             \\ \hline
\multirow{2}{*}{P3}          & NCh05                  & End-user   perspectives are the primary input guiding software creation.                                                                                       & SECO-TransDX is limited to improving the
understanding of DX-driven transparency in SECO.              \\ \cline{2-4} 
                             & NCh06                  & SECO information consumption depends on the structure of the portal.                                                                                        &  This argument is supported by the evidence presented in Parracho et al.~\cite{Parracho2024}.              \\ \hline
\multirow{2}{*}{P4}          & NCh07                  & SECO   portals promote transparency but require complementary conditions, including access, clarity, and governance.                                        &  SECO-TransDX is limited to improving the
understanding of DX-driven transparency in SECO.               \\ \cline{2-4} 
                             & NCh08                  & Transparency in SECO portals varies with ecosystem nature, keystone goals, and available resources.                                                         &   This argument reflects contextual contingencies rather than core elements of the conceptual model.            \\ \hline
P5                           & NCh09                  & Transparency in SECO success depends on multiple influencing conditions.                                                                                     &   This argument is supported by the conditioning factors presented in Zacarias et al.~\cite{Zacarias2024}.             \\ \hline
P6                           & NCh10                  & The analysis of transparency from perspectives may not always be feasible, as not all ecosystems are structured to evaluate transparency comprehensively. &  This argument expresses practical limitations of specific ecosystems, rather than a structural aspect of the conceptual model.              \\ \hline
P8                           & NCh11                  & Transparency benefits DX but its impact varies by developer preferences and context.                                                                        & This argument reflects contextual contingencies rather than core elements of the conceptual model.     \\ \hline
\end{tabular}
\end{table}

Finally, Table~\ref{tab:codesChangesPropositions} shows the comments that suggested refinements to the model. Several focused on clarifying the relationship between SECO portals and common technological platforms (P1--Ch01, P1--Ch03), specifying actor roles (P2--Ch04, P2--Ch06), and acknowledging that developers may also rely on alternative information channels beyond portals (P3--Ch08). Other remarks addressed transparency as a non-functional requirement (P4--Ch09), the somewhat imprecise nature of transparency characteristics (P6--Ch11), and the need for more explicit definitions of DX dimensions and factors (P7--Ch12, P7--Ch13). These refinements were particularly relevant for the subsequent iteration of the conceptual model. Overall, the qualitative analysis complements the statistical findings by showing that most propositions were perceived as consistent, while also identifying areas where additional clarity or refinement was beneficial.

\begin{table}[!ht]
\footnotesize
\caption{Codes associated with comments on SECO-TransDX propositions that change (Ch) the conceptual model in the first round.}
\label{tab:codesChangesPropositions}   
\begin{tabular}{ccp{7cm}p{5cm}}
\hline
\multicolumn{1}{l}{Proposition} & \multicolumn{1}{l}{ID} & Code                                                                                                               & Action \\ \hline
\multirow{3}{*}{P1}             & Ch01                   & SECO portals may offer one or more artifacts.                                                                      &   Adjust the multiplicity.     \\ \cline{2-4} 
                                & Ch02                   & SECO portals allow the exchange of information and products from a common technological platform.                  &  Adjust P1 to explicitly represent portals as enablers of information and product exchange within common technological platforms.        \\ \cline{2-4} 
                                & Ch03                   & SECO portals are associated with a common technological platform.                                                  &  Adjust the relationship between portal and common technological platform.       \\ \hline
\multirow{3}{*}{P2}             & Ch04                   & Third-party developers develop and consume artifacts and tools in SECO.                                            &  Adjust P2 to clarify the types of third-party developers' interaction with artifacts and tools in SECO.        \\ \cline{2-4} 
                                & Ch05                   & Actors are related not only to the ecosystem as a whole but also to the common technological platforms.            & Clarify the scope of actor relationships to include the common technological platforms.       \\ \cline{2-4} 
                                & Ch06                   & Roles are associated with an actor.                                                                                &   Adjust P2 to explicitly define the association between roles and actors.     \\ \hline
\multirow{2}{*}{P3}             & Ch07                   & Third-party developers consume information from SECO portals to make contributions.                                &   Adjust P3 to emphasize the consumption of portal information as input for contributions.     \\ \cline{2-4} 
                                & Ch08                   & Third-party developers also consume SECO information from alternative channels to make contributions.              & Adjust P3 to include alternative channels of SECO information consumption.       \\ \hline
P4                              & Ch09                   & Transparency is a non-functional requirement of common technological platforms.                                     &   Adjust P4 to present transparency as a non-functional requirement of platforms.     \\ \hline
P5                              & Ch10                   & Transparency in common technological platforms may drive one or more success factors in SECO.                      &  Adjust P5 to represent transparency as a potential driver of multiple success factors and adjust the multiplicity.      \\ \hline
P6                              & Ch11                   & The definition of transparency characteristics is fuzzy.                                                            &   Adjust P6, removing the characteristic constructs, since they are already covered by the conditioning factors.  \\ \hline
\multirow{2}{*}{P7}             & Ch12                   & The definition of DX dimensions is not clear in this context.                                                      &  Adjust P7, removing the dimensions constructs, since they are already covered by the DX factors.      \\ \cline{2-4} 
                                & Ch13                   & DX is influenced by one or more factors in SECO.                                                                   &   Adjust P7 to reflect multiple influencing factors on DX and adjust the multiplicity.     \\ \hline
P8                              & Ch14                   & Transparency affects the DX of third-party developers during their contributions to a common technological platform. &  Adjust P8 to specify the effect of transparency on DX in the contribution process.      \\ \hline
\end{tabular}
\end{table}

\begin{figure}[!ht]
\includegraphics[width=1.0\linewidth]{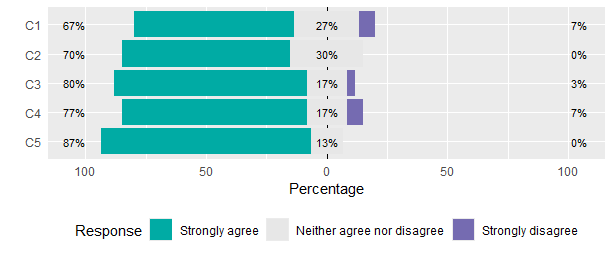}
\caption{Frequency of results for SECO-TransDX criteria evaluation in the first round.}
\label{fig:firstRoundCriteria}
\Description{Graph of frequency of results for SECO-TransDX criteria evaluation in the first round.}
\end{figure}

\paragraph{Descriptive analysis of overall evaluation of SECO-TransDX.}
Table~\ref{tab:firstRoundCriteria} presents the expert panel’s evaluation results for the questions regarding the overall evaluation of SECO-TransDX in the first round, addressing Delphi-RQ2: \textit{``Do the experts agree that SECO-TransDX meets the criteria of ambiguity, explanatory power, parsimony, generality, and utility?''}. The table reports the median, mode, SD, IQR, and percentages of agreement and disagreement. Medians and modes were \textbf{5} for all criteria, with a predominance of agreement, highest in \textbf{C5} (\(86.7\%\)), followed by \textbf{C3} (\(80.0\%\)), \textbf{C4} (\(76.7\%\)), \textbf{C2} (\(70.0\%\)), and \textbf{C1} (\(66.7\%\)). Neutral responses were relatively more frequent in \textbf{C1--C2} (\(\approx 27\%{-}30\%\)) than in \textbf{C3--C5} (\(\approx 13\%{-}17\%\)), and strong disagreement appeared only in \textbf{C1} and \textbf{C4} (\(6.7\%\)) and in \textbf{C3} (\(3.3\%\)), being absent in \textbf{C2} and \textbf{C5}. According to our a priori consensus rule (\(\mathrm{IQR} \le 1\) and \(\mathrm{SD} \le 1.5\)), \textbf{C3--C5} achieved consensus (IQR \(= 0\); SD \(\approx 0.69{-}1.19\)), whereas \textbf{C1--C2} did not (IQR \(= 2\)) despite their high central tendency. Fig.~\ref{fig:firstRoundCriteria} visualizes these distributions, highlighting the ceiling toward positive ratings and the comparatively broader spread for \textbf{C1--C2}.

\begin{table}[!ht]
\footnotesize
\caption{First-round results for SECO-TransDX overall evaluation criteria.}
\label{tab:firstRoundCriteria}  
\begin{tabular}{lllllllll}
\hline
Criteria & Median & Mode & SD & IQR & \% Agreement & Disagreement\\\hline
                        C1     &          5               &     5                  &      1.24               &               2       &    66.7\%                 &  6.7\%     \\
                        C2     &          5               &     5                  &      0.93              &              2        &    70\%                 &  0\%     \\
                        C3     &          5               &      5                 &         1.01            &              0        &    80\%                 &  3.3\%     \\
                        C4     &            5             &      5                 &       1.19              &               0       &    76.7\%                 &  6.7\%     \\
                        C5     &         5                &     5                  &       0.69                &             0         &    86.7\%                 &  0\%     \\
                        \hline
\end{tabular}
\end{table}

In addition to the propositions, experts also provided comments on the evaluation criteria of SECO-TransDX. These insights were similarly coded and grouped into categories, distinguishing between suggestions that only offered contextual clarifications (NCh) and those that pointed to the need for refinements (Ch). 

Table~\ref{tab:codesNoChangesCriteria} shows the codes that did not imply changes to the conceptual model. These comments highlighted, for example, the breadth of certain concepts (C1), the fact that DX extends beyond interaction to include expectations and perceptions (C1), and the need to clarify attribute groupings across criteria (C2, C3). While these remarks did not alter the structure of SECO-TransDX, they added nuance and helped refine the interpretation of its components.

\begin{table}[!ht]
\footnotesize
\caption{Codes associated with comments on overall evaluation criteria that do not change (NCh) the conceptual model in the first round.}
\label{tab:codesNoChangesCriteria}   
\begin{tabular}{ccp{7cm}p{5cm}}
\hline
\multicolumn{1}{l}{Criteria} & \multicolumn{1}{l}{ID} & Code                                                                                                    & Justification \\ \hline
C1                           & NCh12                  & Broad concepts without further context.                                                               & The observation is too general and does not provide actionable input for refining the model.              \\ \hline
C1                           & NCh13                  & DX don't occur only on interactions and cover the expectations and the perceptions of interactions. &   The observation reinforces an already acknowledged view and does not require structural changes.             \\ \hline
C2, C3                       & NCh14                  & Set of attributes on each class.                                                                        &   We aim to provide an understanding of the relationships between constructs rather than an attribute-level detail.    \\ \hline
\end{tabular}
\end{table}

Table~\ref{tab:codesChangesCriteria} presents the codes that led to changes in the conceptual model. These included the need to better specify the relationships between products, developers, and end-users (C1), to revise the representation of transparency characteristics and DX dimensions (C1, C2), and to improve the clarity of the Delphi scenarios (C4) and model usage (C5). In several cases, the experts’ feedback reinforced earlier remarks on the propositions, indicating the convergence of adjustments across both levels (Ch15).

\begin{table}[!ht]
\footnotesize
\caption{Codes associated with comments on overall evaluation criteria that change (Ch) the conceptual model in the first round.}
\label{tab:codesChangesCriteria}   
\begin{tabular}{ccp{7cm}p{5cm}}
\hline
\multicolumn{1}{l}{Criteria} & \multicolumn{1}{l}{ID} & Code                                                                     & Action \\ \hline
C1, C2, C3, C5               & Ch15                   & Improvements based on previous comments on the propositions.             & Align criteria descriptions with the revised propositions to ensure consistency. \\ \hline
C1                           & Ch16                   & Relationships between products, developers, and end-users.               & Clarify the links between products, developers, and end-users. \\ \hline
C1, C2                       & Ch17                   & Revised representation of transparency characteristics and DX dimensions & Update the mapping of transparency characteristics and DX dimensions, since they are already covered by other constructs. \\ \hline
C4                           & Ch18                   & The definition of scenarios in the Delphi questionnaire is not clear.    & Refine the description of scenarios in the question related to C4 to improve clarity and contextual understanding. \\ \hline
C5                           & Ch19                   & The use of the model is not clear.                                       & Change the question related to C5 in the Delphi
questionnaire, adding explicitly the model purpose. \\ \hline
\end{tabular}
\end{table}

\paragraph{Round conclusions.} In Round 1, panel responses generally indicated agreement with the conceptual model. All eight propositions (P1--P8) satisfied the pre-specified item-level consensus criterion 
(\(\mathrm{IQR} \le 1\) and \(\mathrm{SD} \le 1.5\); see Table~\ref{tab:firstRoundPropositions}). At the same time, the qualitative analysis of open comments suggested some refinements to improve clarity in a few propositions, which we carried forward to the next round. For the overall evaluation criteria, consensus was observed for C3--C5, whereas C1--C2 did not meet the stopping condition (\(\mathrm{IQR}=2\)), indicating that additional clarification may be helpful (Table~\ref{tab:firstRoundCriteria}). Guided by these observations, we concluded that our research requires a new Delphi round, revisiting the criteria and incorporating targeted edits informed by the experts’ feedback.

\subsection{Second round}\label{sec:seconRound}
The second round of the Delphi study was conducted between March and April 2025. This round aimed to clarify and refine the results obtained during the first round. To this end, we performed the following actions: (i) we revised the SECO-TransDX conceptual model according to the experts’ recommendations (see Tables\ref{tab:codesChangesPropositions} and \ref{tab:codesChangesCriteria} for details); (ii) we reformulated some propositions to reflect the adjustments made; (iii) we updated the glossary to ensure consistent terminology; and (iv) we modified the Delphi questionnaire to incorporate these refinements. The updated diagram of the SECO-TransDX model, the reformulated propositions, and the revised glossary are provided in the supplementary material at \url{https://doi.org/10.5281/zenodo.16898933}.

To initiate the second round, we sent the new version of the questionnaire along with an anonymous report containing the aggregated results of the first round and all comments made by the experts. After reviewing the report and analyzing the reformulated propositions, experts were invited to either maintain or revise their previous responses. A total of 28 out of 30 experts participated in this round. For those who did not participate, their evaluations from the first round were retained, following the strategy proposed by Torrecilla-Salinas et al.~\cite{Torrecilla-Salinas2019}.

\begin{figure}[!ht]
\includegraphics[width=1.0\linewidth]{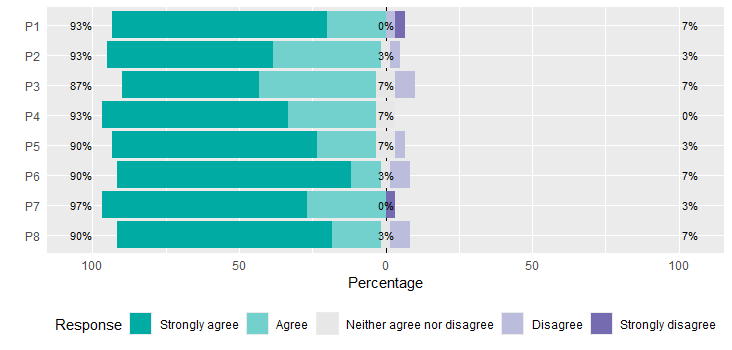}
\caption{Frequency of results for SECO-TransDX propositions evaluation in the second round.}
\label{fig:secondRoundPropositions} 
\Description{Graph.}
\end{figure}

\paragraph{Descriptive analysis of propositions.}
Table~\ref{tab:secondRoundPropositions} summarizes Round~2 results,
again addressing Delphi-RQ1: \textit{``Do the experts agree with the propositions presented for SECO-TransDX?''}. In Round~2, \textbf{all eight propositions met the pre-specified consensus criterion} (\(\mathrm{IQR}\le 1\) and \(\mathrm{SD}\le 1.5\)): medians were high (P1, P2, P4--P8 = \(5.0\); P3 = \(4.0\)), modes were \(5\) for all items, IQR ranged from \(0\) to \(1\), and SD from \(\approx 0.63\) to \(0.94\). Strong disagreement was \textit{marginal}, appearing only in \textbf{P1} and \textbf{P7} (\(3.3\%\) each); overall disagreement remained low (0--6.7\%), and neutrality was limited (0--7\%). Overall agreement (\textit{Strongly agree} + \textit{Agree}) stayed consistently high across items, ranging from \(\approx 86.7\%\) (P3) to \(\approx 96.7\%\) (P7). Fig.~\ref{fig:secondRoundPropositions} illustrates these distributions, confirming the predominance of agreement and a ceiling tendency toward positive ratings, with a slightly broader spread for \textbf{P3} relative to the other propositions.

\begin{table}[!ht]
\footnotesize
\caption{Second-round results for SECO-TransDX propositions.}
\label{tab:secondRoundPropositions}  
\begin{tabular}{lllllllll}
\hline
\multirow{2}{*}{Proposition} & \multirow{2}{*}{Median} & \multirow{2}{*}{Mode} & \multirow{2}{*}{SD} & \multirow{2}{*}{IQR} & \multicolumn{2}{l}{\% Agreement} & \multicolumn{2}{l}{\% Disagreement} \\ \cline{6-9} 
                             &                         &                       &                     &                      & Strongly agree      & Agree      & Disagree     & Strongly disagree    \\ \hline
                        P1     &      5.0                   &   5                    &       0.94              &     0.75                 &     73.3\%                 &  20\%          &    3.3\%                   &      3.3\%       \\
                        P2     &    5.0                      &    5                   &         0.73            &     1                 &      56.7\%                  &     36.7\%       &     3.3\%                  &     0\%        \\
                        P3     &   4.0                        &  5                     &      0.87               &     1                 &      46.7\%               &   40\%         &   6.7\%                    &   0\%          \\
                        P4     &    5.0                       &   5                    &       0.63               &    1                  &      63.3\%               &     30\%       &     0\%                  &      0\%       \\
                        P5     &    5.0                      &    5                   &      0.77               &      1                &      70\%               &     20\%       &        3.3\%               &     0\%        \\
                        P6     &     5.0                     &    5                   &       0.85              &     0                 &      80\%                &   10\%         &       6.7\%                &    0\%         \\
                        P7     &   5.0                       &   5                    &        0.81              &    0.75                  &      70\%               &    26.7\%        &       0\%                &    3.3\%         \\
                        P8     &    5.0                      &    5                   &       0.86              &   1                   &      73.3\%                &    16.7\%        &      6.7\%                 &      0\%       \\ \hline
\end{tabular}
\end{table}

In the second round, the number of comments from experts decreased. Moreover, several of them were related to codes ``reinforce of proposition'' or ``no changes in SECO-TransDX'' previously defined in the first round as presented in the supplementary material at \url{https://doi.org/10.5281/zenodo.16898933}. Hence, we present in this round only codes related to ``changes in SECO-TransDX'' (Table~\ref{tab:codesChangesPropositionsSecondRound}).

\begin{table}[!ht]
\footnotesize
\caption{Codes associated with comments on SECO-TransDX propositions that change (Ch) the conceptual model in the second round.}
\label{tab:codesChangesPropositionsSecondRound}   
\begin{tabular}{ccp{7cm}p{5cm}}
\hline
\multicolumn{1}{l}{Proposition} & \multicolumn{1}{l}{ID} & Code                                                                                                      & Action \\ \hline
P1                              & Ch01                   & The portal should be modeled as supporting or enabling exchanges with the common technological platform & Adjust P1 to explicitly model the portal as a support/enabler of exchanges with the common technological platform. \\ \hline
\multirow{2}{*}{P2}             & Ch02                   & Expand the proposition to include other ecosystem roles beyond third-party developers.                  & Broaden P2 to incorporate additional ecosystem roles beyond third-party developers. \\ \cline{2-4} 
                                & Ch03                   & Actors are associated to common technological platforms.                                                & Adjust P2 to reinforce the association between actors and common technological platforms. \\ \hline
P3, P6                          & Ch04                   & The term ``SECO processes'' is vague and could be better defined.                                       & Revise the terminology of ``SECO processes'' to provide clearer definitions and scope. \\ \hline
\multirow{3}{*}{P3}             & Ch05                   & Relationship between ``contribution'' and ``SECO information''.                                         & Clarify the relationship between contributions and SECO information in P3. \\ \cline{2-4} 
                                & Ch06                   & Multiplicity zero in SECO portals and alternative information channels.                                 & Adjust the multiplicity to allow zero occurrences of SECO portals and alternative channels. \\ \cline{2-4} 
                                & Ch07                   & Multiplicity one or more in SECO information and common technological platform.                         & Adjust the multiplicity to ensure SECO information is linked to one or more common technological platforms. \\ \hline
\multirow{2}{*}{P5}             & Ch08                   & Multiplicity between transparency and common technological platform.                                    & Adjust the multiplicity to represent transparency in relation to the common technological platform. \\ \cline{2-4} 
                                & Ch09                   & Transparency drives at least one success factor.                                                        & Adjust P5 to explicitly state that transparency drives at least one success factor. \\ \hline
P8                              & Ch10                   & Transparency affects the quality of common SECO procedures.                                             & Adjust P8 to include the effect of transparency on the quality of common SECO procedures. \\ \hline
\end{tabular}
\end{table}

\begin{figure}[!ht]
\includegraphics[width=1.0\linewidth]{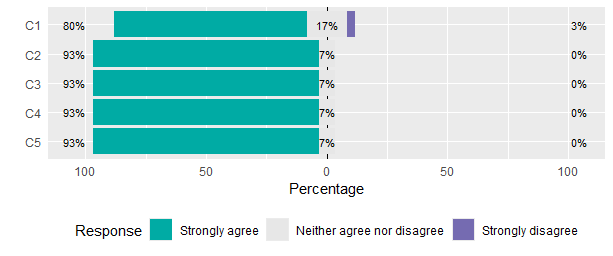}
\caption{Frequency of results for SECO-TransDX criteria evaluation in the second round.}
\label{fig:secondRoundCriteria}
\Description{Graph.}
\end{figure}

\paragraph{Descriptive analysis of overall evaluation of SECO-TransDX.}
Table~\ref{tab:secondRoundCriteria} summarizes Round~2 results, addressing again Delphi-RQ2: \textit{``Do the experts agree that SECO-TransDX meets the criteria of ambiguity, explanatory power, parsimony, generality, and utility?''}. Medians and modes were \textbf{5} for all criteria, with \(\mathrm{IQR}=0\) across the board and \(\mathrm{SD}\approx 0.51{-}1.01\). Agreement predominated, reaching \(93.3\%\) for \textbf{C2--C5} and \(80.0\%\) for \textbf{C1}; neutrality was limited (\(\approx 7\%\) for \textbf{C2--C5} and \(17\%\) for \textbf{C1}), and strong disagreement appeared only marginally in \textbf{C1} (\(3.3\%\)). According to the pre-specified consensus rule (\(\mathrm{IQR}\le 1\) and \(\mathrm{SD}\le 1.5\)), \textbf{all five criteria achieved consensus} in Round~2. Fig.~\ref{fig:secondRoundCriteria} visualizes these distributions, confirming the ceiling tendency toward positive ratings and the slightly broader spread observed for \textbf{C1} relative to the other criteria.

\begin{table}[!ht]
\footnotesize
\caption{Second-round results for SECO-TransDX overall evaluation criteria.}
\label{tab:secondRoundCriteria}  
\begin{tabular}{lllllllll}
\hline
Criteria & Median & Mode & SD & IQR & \% Agreement & Disagreement\\\hline
 C1     &          5               &     5                  &      1.01               &               0       &    80\%                 &  3.3\%     \\
 C2     &          5               &     5                  &      0.51              &              0        &    93.3\%                 &  0\%     \\
 C3     &          5               &      5                 &         0.51            &              0        &    93.3\%                 &  0\%     \\
 C4     &          5             &      5                 &       0.51              &               0       &    93.3\%                 &  0\%     \\
 C5     &         5                &     5                  &       0.51                &             0         &    93.3\%                 &  0\%     \\
                        \hline
\end{tabular}
\end{table}

\paragraph{Round conclusions.}
As a main outcome of Round~2, the uncertainties raised in Round~1 were largely addressed through targeted refinements to wording and scope. The descriptive analyses indicate that the expert panel reached item-level consensus for both the propositions (P1--P8) and the overall evaluation criteria (C1--C5), according to the pre-specified stopping condition (\(\mathrm{IQR}\le 1\) and \(\mathrm{SD}\le 1.5\); see Tables~\ref{tab:secondRoundPropositions} and \ref{tab:secondRoundCriteria}). Agreement remained high across items, and the small shifts observed between rounds are consistent with the incorporation of clarifications. Overall, Round~2 supports the refined SECO-TransDX propositions and the criteria set proposed by Sjøberg et al.~\cite{Sjøberg2008}. With item-level consensus reached, we concluded the Delphi step of this work.

\subsection{Final version of the conceptual model}\label{sec:finalVersion}
\begin{figure}
\includegraphics[width=0.95\linewidth]{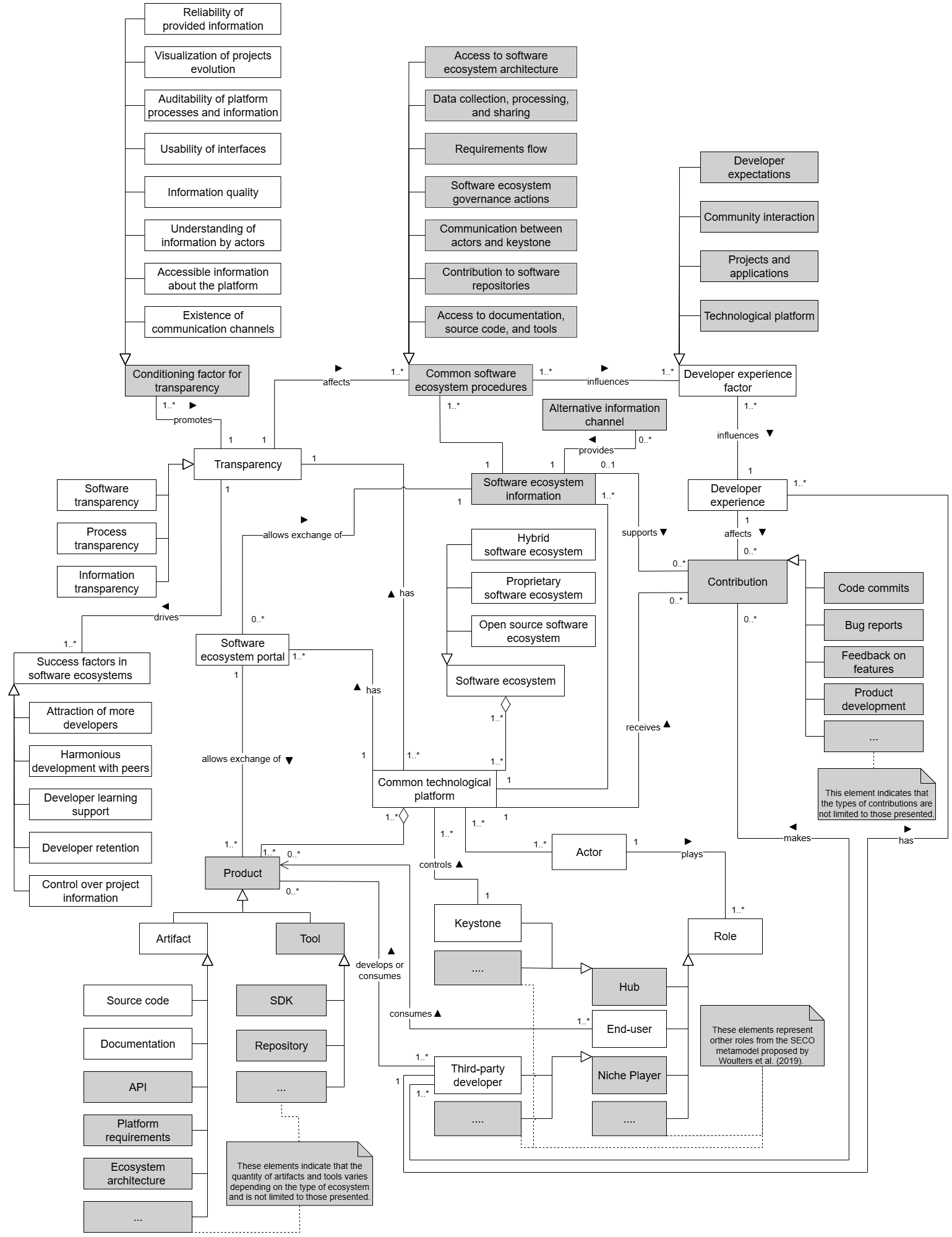}
\caption{Final version of the conceptual model for Transparency in SECO from a DX Perspective with refinements highlighted in gray after the second round.}
\label{fig:SECO-TransDX_final} 
\Description{A conceptual model design with UML notation.}
\end{figure}

As presented earlier, our research was guided by the following RQ: 
\textit{``How can transparency in SECO be characterized from a DX perspective?''} To answer it, Fig.~\ref{fig:SECO-TransDX_final} shows the diagrammatic representation of the final version of SECO-TransDX, comprising 63 concepts and their relationships. The figure also highlights (in gray) the modifications made after two Delphi evaluation rounds. Table~\ref{tab:propositions_final} presents the final versions of the propositions, with the changes highlighted in italics, and the updated glossary is available in the supplementary material at \url{https://doi.org/10.5281/zenodo.16898933}.

By consolidating the conceptual model, its propositions, and glossary, the final version of SECO-TransDX addresses this question. It provides a structured conceptualization of transparency from a DX perspective, clarifying how transparency can be represented and analyzed within SECO.

\begin{table}[!ht]
\scriptsize
\caption{Final version of the statement of SECO-TransDX propositions}
\label{tab:propositions_final}      
\begin{tabular}{p{0.03cm} p{6cm} p{8cm}}
\hline\noalign{\smallskip}
ID & Proposition & Explanation  \\
\noalign{\smallskip}\hline\noalign{\smallskip}

P1 & \underline{Software ecosystem portals} \textit{allow the exchange of \underline{information} and \underline{products} (\underline{artifacts} and \underline{tools})} from a \underline{common technological platform}, considering the type of \underline{software ecosystem} \textit{(\underline{open source}, \underline{proprietary}, or \underline{hybrid})}. & \textit{SECO portals allow the exchange of information and one or more products from a common technological platform. These products can be artifacts and tools. Possible types of artifacts are (not limited to): source code, documentation, API, platform requirements, SECO architecture. Possible types of tools are (not limited to): SDK and repository~\cite{Jansen2009, meireles2019instrumento, Wouters2019}.} The \textit{products} available on the portals vary according to the type of SECO, which can be classified as open source, proprietary, or hybrid~\cite{Manikas2016}. \\ \hline

P2 & \textit{A \underline{third-party developer} is one of the \underline{roles} of \underline{actor}s in \underline{software ecosystems} who develop or consume zero or many \underline{products} of} a \underline{common technological platform}, which is controlled by a \underline{keystone}, \textit{for consumption by one or many \underline{end-users}}. & SECO are based on the collaboration of multiple actors~\cite{Manikas2016}. \textit{According to Wouters et al.~\cite{Wouters2019}, we can identify (not limited to) some main actors’ roles in SECO: hub, niche player, keystone, third-party developer, among others. A third-party developer is one of the roles of actors who may develop or consume zero or many products of one or more common technological platforms for consumption by one or many end-users~\cite{Hanssen2012, Wouters2019}. A keystone is an organization or group that drives the development of one or more common technological platforms~\cite{Wouters2019, Lewellen2020}}. \\ \hline

P3 & \textit{A \underline{third-party developer} makes \underline{contributions} (e.g. code commits, bug reports, feedback on features, product development etc.) to a \underline{common technological platform} with the support of \underline{software ecosystem information} provided by \underline{software ecosystem portals} or \underline{alternative information channels} (e.g. community forums, blogs, or external tutorials etc.) regarding \underline{common software ecosystems procedures}}. & SECO portals are the main source of information related to the products that constitute a common technological platform~\cite{meireles2019instrumento}. \textit{Therefore, a third-party developer makes contributions to a common technological platform with the support of SECO information provided by SECO portals or alternative information channels (e.g., community forums, blogs, external tutorials etc.) regarding common SECO procedures. So that, they can be aware of the processes and elements that are part of such a platform and necessary} to make their contributions~\cite{Souza2020, Parracho2024}. \textit{It is important to highlight that these developers can use information from alternative channels (e.g., community forums, blogs, external tutorials etc.) to support their contributions~\cite{Parracho2024}}. 
 \\ \hline

P4 & \textit{\underline{Transparency} is a non-functional requirement of \underline{common technological platforms} and is promoted by one or more \underline{conditioning factors} in \underline{software ecosystems} (i.e., \underline{existence of communication channels}, \underline{accessible information about the platform}, \underline{understanding of} \underline{information by actors}, \underline{information quality}, \underline{usability of interfaces}, \underline{auditability of platform processes and information}, \underline{visualization of}  \underline{projects evolution}, and \underline{reliability of the provided information})}.  & In SECO, transparency is considered from the software perspective, i.e., a condition that encompasses openness, clarity, and visibility of the mechanisms, processes, and actions of software applications~\cite{LeiteCappelli2010, Isong2022}. \textit{It is a non-functional requirement of one or more common technological platforms and allows actors to access its information and common procedures and learn how to make contributions to it~\cite{Cataldo2010, SantosEtAl2016}. Transparency is supported by one or more conditioning factors in SECO, which are defined as elements, characteristics, or actions that are necessary but not sufficient for the transparency of SECO~\cite{Zacarias2024}. These factors are:} existence of communication channels, accessible information about the platform, understanding of information by actors, information quality, usability of interfaces, auditability of platform processes and information, visualization of projects evolution, and reliability of the provided information~\cite{Linåker2016, Souza2020, RUNESON2021111088, BEELEN2022102733}. \\ \hline

P5 & \textit{\underline{Transparency} in \underline{common technological platforms} drives one or more \underline{success factors in software ecosystems}, such as \underline{attracting more developers}, \underline{harmonious development with peers}, \underline{supporting developer learning of market mechanisms}, \underline{developer retention}, and \underline{control over project information}}. & \textit{Transparency in common technological platforms drives one or more success factors in SECO, i.e., positive advantages of promoting transparency in SECO~\cite{Cataldo2010, Zacarias2024}.} Some examples of success factors are attracting more developers~\cite{BEELEN2022102733}, harmonious development with peers~\cite{Zacarias2024}, supporting developer learning of market mechanisms~\cite{Dabbish2012}, developer retention~\cite{meireles2019instrumento}, and control over project information~\cite{Linåker2016}. The success factors contribute to the sustainability of SECO~\cite{fontao2021developer}. \\ \hline

P6 & \underline{Transparency} can be analyzed in \underline{software ecosystems} from \underline{information}, \underline{process}, and \underline{software perspectives} \textit{to understand its effects on the quality of one or more \underline{common software ecosystems procedures}, such as \underline{access to} \underline{documentation, source code, and tools}; \underline{contribution to software}  \underline{repositories}; \underline{communication between actors and keystone}; \underline{software ecosystem governance actions}; \underline{requirements flow}; \underline{data collection, processing, and sharing}; and \underline{access to} \underline{ecosystem architecture}}. & Considering related needs in the software context, which includes SECO, Leite and Cappelli~\cite{LeiteCappelli2010} \textit{state that transparency is a non-functional requirement and must be considered at all stages of software design to clarify how software provides transparency in organizational processes and information~\cite{Cysneiros2013, Hosseini2015}. Therefore,} transparency can be analyzed in SECO from information, process, and software perspectives~\cite{Parracho2024} \textit{to understand its effects the quality of one or more common SECO procedures, such as access to documentation, source code, and tools; contribution to software repositories; communication between actors and keystone; SECO governance actions; requirements flow; data collection, processing, and sharing; and access to ecosystem architecture. The levels of transparency for these processes depended on the business context and organizational objectives of each SECO~\cite{Zacarias2024}}. \\ \hline

P7 & \textit{\underline{Developer experience} in \underline{software ecosystems} is influenced by one or more \underline{factors} related to \underline{technological platform}, \underline{projects and applications}, \underline{community interaction}, and \underline{developer expectations}.} & According to Fagerholm and Münch~\cite{Fagerholm2012}, \textit{DX consists of experiences relating to all kinds of artifacts and activities that a developer may encounter as part of their involvement in software development.} In SECO, some specific factors influence DX based on the characteristics related to the dynamics of SECO: (a) factors of technological platform, which are related to the technical infrastructure for development provided by a common technological platform, \textit{e.g. desired technical resources for development,  easy to configure platform etc.~\cite{Kauschinger2021, STEGLICH2023111808}}; (b) factors of projects and applications, which are related to the process of developing and distributing applications on a common technological platform, \textit{e.g. more clients/users for applications, application distribution methods etc.~\cite{Choia2017Impacts, STEGLICH2023111808}}; (c) factors of community interaction, which are related to the interaction between a developer and other developers who integrate a SECO community, \textit{e.g. obtaining community recognition, commitment to the community etc.~\cite{Fontao2020, STEGLICH2023111808}}; and (d) factors of developer expectations, which are related to expectations and benefits obtained by a developer’s contribution in interacting with a SECO, \textit{e.g. emergence of new market and job opportunities, more financial gains, etc.~\cite{DeSouza2016, STEGLICH2023111808}}.   \\ \hline

P8 & \textit{\underline{Transparency} affects the quality of one or more \underline{common software ecosystem procedures} that influence one or more \underline{factors} of the \underline{developer experience} of \underline{third-party developers} during their \underline{contributions} to a \underline{common technological platform}}. & Third-party developers access SECO portals \textit{or alternative channels to consume information about common SECO procedures to learn how to contribute to a common technological platform}~\cite{meireles2019instrumento, Parracho2024}. \textit{The quality of one or more of these processes influences one or more factors of the DX of third-party developers during their contributions. When the quality of these processes is not sufficient,} these third-party developers may have difficulty developing their applications autonomously. This situation is related to the level of transparency of these processes~\cite{meireles2019instrumento}. \textit{Low or} lack of transparency contributes to increased barriers to entry for a SECO and a cumbersome start for newcomer third-party developers. This issue can negatively influence one or more DX factors and cause software developers to give up contributing to a common technological platform~\cite{SantosEtAl2016, knauss2018continuous, meireles2019instrumento, fontao2021developer}. \\
\noalign{\smallskip}\hline
\end{tabular}
\end{table}

\section{Discussion}\label{sec:discussion}
This section depicts our contributions and main findings, and future perspectives on the use of SECO-TransDX.

\subsection{Contributions and main findings}\label{sec:mainFindings}
After two Delphi rounds with domain experts, the final version of SECO-TransDX consolidates a developer-centered conceptualization of transparency in SECO. The iterative process combined quantitative consensus measures with qualitative feedback to progressively clarify constructs and relationships, resulting in the final set of propositions (P1--P8) and their diagrammatic representation (Fig.~\ref{fig:SECO-TransDX_final} and Table~\ref{tab:propositions_final}). The refinements emphasized the role of SECO portals as enablers of information and product exchange, the association between actors and common technological platforms, and the linkage between contributions and ecosystem information, while maintaining generality across proprietary, open-source, and hybrid SECO. In this consolidated form, SECO-TransDX offers a clear and traceable account of \textit{DX-driven transparency} that aligns conceptual definitions, relationships, and scope.

The consolidation of SECO-TransDX contributes to bridging the theoretical gap outlined in Section~\ref{sec:relatedWork}. As discussed, prior research often treats transparency and DX as separate constructs: transparency primarily as an organizational or governance attribute~\cite{Oliveira2020}, and DX as a cognitive and motivational phenomenon~\cite{Fagerholm2012, Greiler2022}. This separation limits the understanding of how technical transparency mechanisms shape developers’ day-to-day experiences and engagement within SECO.  

\paragraph{Theoretical Contribution.} SECO-TransDX advances this discussion by explicitly integrating transparency and DX into a unified conceptual model. Building on Wouters et al.~\cite{Wouters2019}, who established a shared vocabulary for SECO concepts but did not address how developers experience transparency, our model positions transparency as an experiential construct directly linked to DX. Likewise, while Oliveira et al.~\cite{Oliveira2020} emphasize transparency from a governance perspective, SECO-TransDX reframes it through a developer-centered lens, showing how transparency mechanisms condition perceptions of fairness, clarity, and usability. In line with Malcher et al.~\cite{MALCHER2025107672}, who highlight coordination and requirements management challenges, our model complements their findings by demonstrating how transparency mediates these processes and shapes long-term developer satisfaction and retention.  

In this way, SECO-TransDX introduces the notion of \textit{DX-driven transparency in SECO}, offering a novel theoretical lens for SECO research. Rather than conceptualizing transparency only as a static organizational feature, the model articulates its dynamic role in shaping developers’ perceptions, motivations, and participation. This integration of transparency and DX extends prior conceptual models and provides a structured basis for analyzing how ecosystem practices affect developer engagement and, ultimately, ecosystem sustainability.

\paragraph{Practical Contribution.} Beyond its theoretical role, SECO-TransDX provides a practical foundation for guiding keystones and portal managers in structuring transparency practices with a clear focus on DX. By articulating how the conditioning factors for transparency directly shape perceptions of fairness, clarity, and usability, the model offers concrete elements that can be operationalized in practice.  

First, the model can support \textit{portal design and improvement}. For instance, ensuring accessible and regularly updated documentation (linked to P1 and P2) or maintaining clear governance disclosures (P4) can be prioritized not only as organizational necessities but also as developer-centered transparency practices. In this way, SECO-TransDX complements existing operational guidelines by making explicit the connection between transparency features and the quality of the developer journey.  

Second, the model can serve as a \textit{diagnostic and evaluation tool}. Keystones may use SECO-TransDX as a reference for assessing current transparency practices, identifying gaps, and prioritizing improvements with a direct impact on developer satisfaction and retention. For example, by mapping existing portal functionalities against the propositions (P1--P8), managers can detect where transparency mechanisms may be insufficient or fragmented, and thus risk undermining developer trust.  

Finally, SECO-TransDX has value for \textit{strategic decision-making}. In hybrid or proprietary ecosystems, where transparency is often balanced with control, the model clarifies how specific design choices, such as limiting or opening access to repositories, or formalizing decision-making processes, affect developers’ perceptions and engagement. This makes it possible for ecosystem orchestrators to reason about trade-offs in transparency not only from a governance or business perspective, but also from the lens of DX, aligning sustainability goals with developer needs.  

\paragraph{Main findings} From the two Delphi rounds, five main findings emerged, consolidating how transparency is experienced and operationalized in SECO from a developer-centered perspective. Together, these findings advance the understanding of \textit{DX-driven transparency in SECO}, clarifying how transparency mechanisms are perceived and enacted through developers’ interactions with ecosystem platforms and procedures:

\begin{itemize}
    \item \textbf{Transparency as an interactive experience, not only informational.} Transparency in SECO is not limited to the availability of artifacts such as code, documentation, or metrics. Instead, it also involves how such information is perceived, interpreted, and made actionable by developers during their interaction with the ecosystem. This finding emphasizes the experiential dimension of transparency, complementing the more structural perspectives found in the literature~\cite{Fagerholm2012, Greiler2022, Wouters2019};

    \item \textbf{Governance transparency as a key dimension of DX.} Transparency in governance processes, such as decision-making structures, policies, and communication channels, emerged as essential to how developers evaluate fairness, trust, and long-term engagement within the ecosystem. While prior work has highlighted transparency mainly as an attribute of governance~\cite{Oliveira2020}, SECO-TransDX reframes it through a developer-centered lens, showing how it directly shapes the DX;

    \item \textbf{SECO portals as mediators and enablers of DX-driven transparency.} 
    SECO portals were reinforced as the primary interfaces through which transparency is mediated and experienced by developers, affecting onboarding, collaboration, and retention. They act not only as repositories of information but as socio-technical spaces where transparency becomes actionable. This aligns with the view of SECO as socio-technical systems where artifacts and interactions are closely intertwined~\cite{Zacarias2024, Cataldo2010};

    \item \textbf{Transparency as a crosscutting concern across common ecosystem procedures.} Transparency permeates all SECO procedures related to requirements management, governance, communication, and architecture, conditioning how developers perceive their role and motivation to contribute. This finding complements models that address transparency within a single dimension (e.g., governance~\cite{Oliveira2020} or requirements~\cite{MALCHER2025107672}), highlighting its systemic role across the ecosystem;

    \item \textbf{Transparency as a mediator of DX and sustainability.} 
    Beyond individual processes, transparency was consolidated as a mediating condition that links ecosystem practices with long-term developer satisfaction and participation. In this sense, SECO-TransDX positions transparency as a mechanism that influences not only immediate usability or clarity, but also broader sustainability outcomes in SECO~\cite{Greiler2022, Fagerholm2012, Zacarias2024sesos, zacarias2025Dxfactors}.
\end{itemize}

\subsection{Perspectives for using SECO-TransDX}\label{sec:perspectives}
The consolidation of SECO-TransDX as a conceptual model opens up multiple avenues for future research and practice. Beyond synthesizing transparency and DX into a unified lens, the model provides a flexible foundation that can be mobilized in different ways depending on the objectives of researchers and practitioners. The following perspectives illustrate how SECO-TransDX may be employed: as a basis for empirical validation and case studies, as guidance for portal design and evaluation, as a bridge to Integration with governance and sustainability research, as a basis for transparency guidelines for SECO portals, and as an adaptable framework with potential applications across different domains.

\paragraph{Empirical validation and case studies.} SECO-TransDX can be applied and validated empirically in different SECO contexts. Future studies may use focus groups, controlled experiments, or case studies to examine how the propositions (P1–P8) materialize in practice, and whether their effects on DX vary across proprietary, open-source, and hybrid ecosystems. For instance, evaluating portals such as Android, iOS, or HarmonyOS Next would enable testing how transparency mechanisms concretely affect the developer journey. 

\paragraph{Guidance for portal design and evaluation.} The model also serves as a diagnostic and evaluative framework for SECO portals. By mapping existing transparency mechanisms against SECO-TransDX, keystones and portal managers can identify gaps and prioritize improvements with a direct impact on DX. This perspective highlights the practical value of the model for informing portal design choices and for creating dashboards or assessment tools that visualize transparency practices in terms of developer experience. 

\paragraph{Integration with governance and sustainability research.} SECO-TransDX can be extended to research linking transparency, governance, and sustainability. As transparency mediates developers’ trust and motivation, it provides a new lens to understand how governance decisions (e.g., openness vs. control, or centralized vs. participatory decision-making) shape ecosystem resilience. This integration bridges developer-centered perspectives with long-standing debates in governance and ecosystem sustainability. 

\paragraph{Basis for transparency guidelines for SECO portals.} Another perspective is the use of SECO-TransDX as a foundation for elaborating structured transparency guidelines for SECO portals. By linking conditioning factors and propositions to concrete portal practices, the model provides the conceptual grounding for defining actionable criteria that portal managers can adopt. This makes it possible to transform abstract concepts of DX-driven transparency into prescriptive recommendations, ensuring that transparency practices are consistently aligned with developer needs. 

\paragraph{Cross-domain adoption.} Finally, the notion of DX-driven transparency introduced by SECO-TransDX has potential applications beyond software ecosystems. Its principles may inform research and practice in other domains where openness and developer-oriented participation are central, such as artificial intelligence (AI) ecosystems, data platforms, or open science infrastructures. This opens a path for testing the generalizability of the model in contexts where transparency, usability, and participant engagement are intertwined.

\section{Threats to validity}\label{sec:threats}

As with any research effort that combines conceptual modeling, synthesis of prior evidence, and expert-based evaluation, this work is subject to several threats to validity. We discuss them according to Wohlin et al.’s classification~\cite{Wohlin2012}: construct validity, internal validity, external validity, and reliability, along with the mitigation strategies we adopted.

\subsection{Construct validity}
Construct validity refers to whether the operationalization of constructs and propositions adequately represents the theoretical concepts under investigation~\cite{Wohlin2012}. A main threat in our study was the risk of misrepresenting or omitting relevant concepts of transparency and DX when designing SECO-TransDX. To mitigate this, we built on a solid foundation of prior studies on DX~\cite{zacarias2025Dxfactors} and transparency in SECO~\cite{Zacarias2024}, complemented by the SECO meta-model~\cite{Wouters2019}. Coding and construct identification were conducted iteratively, with multiple review cycles among the authors to ensure consensus and theoretical alignment. During the Delphi evaluation, we piloted the questionnaire to refine its design and reduce ambiguities. Furthermore, the questionnaire items were mapped directly to constructs and propositions, ensuring alignment between theoretical definitions and operational measures.

\subsection{Internal validity}
Internal validity concerns uncontrolled factors that may influence study results. In our work, potential threats included author bias during the synthesis of prior evidence, framing effects in questionnaire design, and experts’ predispositions influencing Delphi responses. To mitigate these, we defined research protocols for study selection, coding, and expert recruitment before data collection. Delphi responses were anonymized to minimize social desirability bias and panel dominance. Moreover, multiple Delphi rounds enabled experts to revise their judgments after considering group feedback, helping reduce individual bias effects.

\subsection{External validity}
External validity refers to the generalizability of the findings beyond the studied context. In our case, threats arise from the limited number of experts and the heterogeneity of SECO contexts. Although our Delphi panel included 30 participants from academia, industry, and open-source communities with varying levels of expertise, it may not fully capture the diversity of SECO stakeholders. To mitigate this, we complemented expert judgment with constructs grounded in prior empirical studies and designed the model to be independent of a specific SECO type (proprietary, open-source, or hybrid). Nevertheless, we acknowledge that further empirical validation through case studies in real ecosystems (e.g., Android, iOS, HarmonyOS) is required to strengthen the generalizability of SECO-TransDX.

\subsection{Reliability}
Reliability refers to the consistency and repeatability of the research process. A threat in our study concerns the subjectivity involved in coding results, synthesizing constructs, and interpreting experts’ comments. To address this, coding was first performed by one author and independently reviewed by the others, with disagreements resolved through discussion until consensus was reached. In the Delphi analysis, we predefined the stopping criteria ($IQR \leq 1$ and $SD \leq 1.5$), ensuring transparency and consistency in determining when consensus was achieved. All methodological steps (from construct identification to Delphi procedures) were carefully documented, increasing traceability and supporting replicability of our study.

\section{Conclusion}\label{sec:conclusion}

The complexity and openness of SECO make transparency both a necessary and challenging property to conceptualize and operationalize. The main purpose of this work was to advance a developer-centered understanding of transparency in SECO, contributing to the consolidation of this emerging area of research. To this end, we proposed \textit{SECO-TransDX}, a conceptual model that organizes how transparency influences DX in SECO, integrating constructs, propositions, and their relationships.

The construction of SECO-TransDX was grounded in the body of knowledge provided by prior secondary studies on DX~\cite{zacarias2025Dxfactors}, transparency in SECO~\cite{Zacarias2024}, and the SECO meta-model~\cite{Wouters2019}. The model identified key concepts, conditioning factors, common ecosystem procedures, artifacts, and relational dynamics that shape how transparency is perceived and constructed from the developer’s perspective during interactions with SECO. The model was evaluated through a Delphi study with experts from academia and industry, leading to refinements and consensus around its propositions. This evaluation process generated five main findings that consolidate the notion of DX-driven transparency and highlight its relevance to both research and practice.

Beyond its theoretical contributions, SECO-TransDX also has implications for both academia and industry. For researchers, it offers a structured lens to analyze how transparency mediates DX and provides a foundation for developing guidelines, evaluation frameworks, and practical tools to support more transparent and developer-centered platforms. For practitioners, the model illustrates how different factors and interactions influence developers’ perception of transparency, providing a basis for designing transparent practices and platforms that pave the way for more sustainable and trustworthy SECO. 

The main takeaway message of this article is that transparency is not only a non-functional requirement in SECO, but also a decisive factor shaping DX. In summary, SECO-TransDX advances the understanding of how transparency can be characterized, evaluated, and applied in SECO. By structuring this knowledge through a developer-centered lens, the model contributes to both theoretical consolidation and practical pathways for fostering more transparent and sustainable software ecosystems.

From the proposition of SECO-TransDX, we envisioned future work from the perspective of the SECO research community, as in the following: (i) the continuous evolution of SECO-TransDX as new evidence on transparency and DX emerges; (ii) empirical validation of the model in different types of SECO (e.g., open-source, proprietary, and hybrid), identifying how transparency manifests across diverse settings; (iii) the application of the model as a conceptual basis for developing guidelines and evaluation frameworks that support platform managers in improving transparency in SECO portals; and (iv) the use of SECO-TransDX as a starting point for designing tools, dashboards, or practices that operationalize transparency in developer-centered ecosystems.

\begin{acks}
This work was financed in part by the Coordenação de Aperfeiçoamento de Pessoal de Nível Superior – Brazil (CAPES) – Finance Code 001 and Grant 88887.928989/2023-00, CNPq (Grant 316510/2023-8), FAPERJ (Grant E-26/204.404/2024), and UNIRIO.
\end{acks}

\bibliographystyle{ACM-Reference-Format}
\bibliography{bibliography}


\end{document}